\documentclass[%
 reprint, 
superscriptaddress,
 amsmath,amssymb,
 aps, physrev,
]{revtex4-2}

\usepackage{graphicx}
\usepackage{dcolumn}
\usepackage{bm}
\usepackage{siunitx}
\usepackage{textgreek}
\usepackage{soul}
\usepackage{tabularx}
\usepackage{xcolor}

\begin{document}

\preprint{APS/123-QED}

\title{\textbf{Krypton-sputtered tantalum films for scalable high-performance quantum devices} 
}%

\author{Maciej W. Olszewski}
\email{mwo34@cornell.edu}
\affiliation{Department of Physics, Cornell University, Ithaca, NY 14853, USA}%

\author{Lingda Kong}
\affiliation{School of Applied and Engineering Physics, Cornell University, Ithaca, NY 14853, USA}%
 
\author{Simon Reinhardt}
\email{sfr64@cornell.edu}
\affiliation{School of Applied and Engineering Physics, Cornell University, Ithaca, NY 14853, USA}%

\author{Daniel Tong}
\affiliation{School of Applied and Engineering Physics, Cornell University, Ithaca, NY 14853, USA}
\affiliation{Department of Materials Science and Engineering, Cornell University, Ithaca, NY 14850, USA}%

\author{Xinyi Du}
\affiliation{Cornell NanoScale Facility, Cornell University, Ithaca, NY 14853, USA}%

\author{Gabriele Di Gianluca}
\affiliation{Department of Physics, University of Florida, Gainesville, FL 32611, USA}%

\author{Haoran Lu}
\affiliation{School of Applied and Engineering Physics, Cornell University, Ithaca, NY 14853, USA}%

\author{Saswata Roy}
\affiliation{Department of Physics, Cornell University, Ithaca, NY 14853, USA}%

\author{Luojia Zhang}
\affiliation{Department of Physics, Cornell University, Ithaca, NY 14853, USA}%

\author{Aleksandra B. Biedron}
\affiliation{NY Creates, Albany, NY 12203, USA}%

\author{David A. Muller}
\affiliation{School of Applied and Engineering Physics, Cornell University, Ithaca, NY 14853, USA}%
\affiliation{Kavli Institute at Cornell for Nanoscale Science, Cornell University, Ithaca, NY 14853, USA}%

\author{Valla Fatemi}
\email{vf82@cornell.edu}
\affiliation{School of Applied and Engineering Physics, Cornell University, Ithaca, NY 14853, USA}%

\date{\today}

\begin{abstract}

Superconducting qubits based on tantalum (Ta) thin films have demonstrated the highest-performing microwave resonators and qubits~\cite{bland_millisecond_2025}.
This makes Ta an attractive material for superconducting quantum computing applications, but, so far, direct deposition has largely relied on high substrate temperatures exceeding \SI{400}{\celsius} to achieve the body-centered cubic phase, BCC (\textalpha-Ta)~\cite{crowley_disentangling_2023,lozano_low-loss_2024}.
This leads to compatibility issues for scalable fabrication leveraging standard semiconductor fabrication lines~\cite{awschalom_challenges_2025}.
Here, we show that changing the sputter gas from argon (Ar) to krypton (Kr) promotes BCC Ta synthesis on silicon (Si) at temperatures as low as \SI{200}{\celsius}, providing a wide process window compatible with back-end-of-the-line fabrication standards. 
Furthermore, we find these films to have substantially higher electronic conductivity, consistent with clean-limit superconductivity.
We validated the microwave performance through coplanar waveguide resonator measurements, finding that films deposited at \SI{250}{\celsius} and \SI{350}{\celsius} exhibit a tight performance distribution at the state of the art.
Higher temperature-grown films exhibit higher losses, in correlation with the degree of Ta/Si intermixing revealed by cross-sectional transmission electron microscopy. 
Finally, with these films, we demonstrate transmon qubits with a relatively compact, \SI{20}{\micro\meter} capacitor gap, achieving a median quality factor up to 14 million. 

\end{abstract}

\maketitle

\section{\label{sec:intro}Introduction}
Materials and processing innovations underpinned major advances in solid state technology~\cite{awschalom_challenges_2025}, and 
superconducting quantum information hardware are experiencing development bottlenecks due to performance issues linked to the underlying materials~\cite{de_leon_materials_2021,mcrae_measurement_2022}.
Academic efforts frequently advance approaches to enhance the performance of fabricated chips, as has recently been promoted by research investigations promoting the body-centered cubic phase (BCC) of tantalum (\textalpha-Ta) as a resilient, high-performing material~\cite{place_new_2021,crowley_disentangling_2023,lozano_low-loss_2024,ganjam_surpassing_2024,marcaud_low-loss_2025,bland_millisecond_2025}.
Detailed, iterative nanofabrication triage and materials characterization~\cite{mcrae_reproducible_2021,torres-castanedo_formation_2024,kono_mechanically_2024,  olszewski_low-loss_2025,gingras_improving_2025,bland_millisecond_2025} has found leading-edge performance when Ta is directly deposited onto high-quality substrates like silicon (Si). 
Thereby, transmon qubits with millisecond coherence have been achieved in academic labs~\cite{bland_millisecond_2025}.

\begin{figure}[t]
\includegraphics[scale = 1]{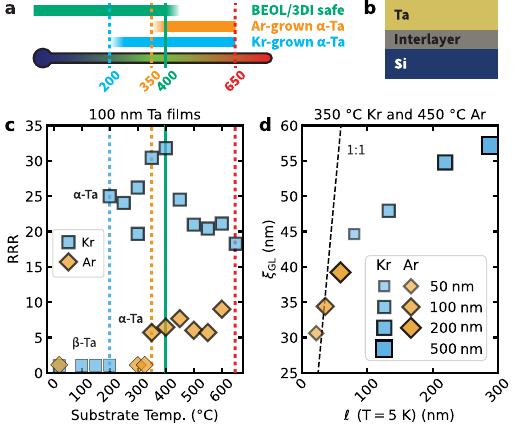}
\caption{\label{fig:intro}
(a) Schematic for the temperature deposition windows of \textalpha-Ta on Si wafers with Ar and Kr sputter gases and their comparison with BEOL process windows.
(b) Schematic of the cross-section profile of the films.
(c) Residual resistivity ratio (RRR) \SI{100}{\nano\meter}-thick \textalpha-Ta films as a function of substrate temperature during deposition for the two process gases. 
(d) The determined Ginzburg-Landau coherence length of superconductivity as a function of the effective transport mean free path ($\ell_{\rm 5K}$) for films of different thickness. The dashed line represents a 1:1 line.}
\end{figure}

The translation of such performance to industrial settings is a critical next step.
It is expected that semiconductor foundries will provide large improvements in reliability and throughput.
The scalability of a method can be understood through the lens of semiconductor chip manufacturing process windows; compatibility with these process windows will substantially lower the barrier to translating fabrication processes to production~\cite{awschalom_challenges_2025}. 
In particular, when metals are involved, back-end-of-the-line (BEOL) processing is the norm, which sets an upper temperature limit of \SI{400}{\celsius} (Fig.~\ref{fig:intro}a).
\textalpha-Ta films are usually deposited with magnetron sputtering using Ar process gas, and this process has typically required substrate temperatures between \SI{450}{\celsius} and \SI{650}{\celsius}~\cite{place_new_2021,lozano_low-loss_2024,crowley_disentangling_2023,zikiy_investigation_2025}, which is in conflict with BEOL processing requirements. 
A scalable approach to depositing \textalpha-Ta directly onto silicon is needed.
This challenge presents an interesting arena for materials synthesis science due to the tension with bulk thermodynamics, in which \textalpha-Ta is the only stable phase of pure Ta.

Here, we demonstrate that using Kr as a sputter gas results in \textalpha-Ta thin films on Si(100) wafers at substrate temperatures as low as \SI{200}{\celsius}, providing a wide BEOL-compatible process window.
We further find the Kr-deposited films to have substantially higher electronic conductivity than Ar-deposited films, entering the clean limit of superconductivity.
Transmission electron microscopy (TEM) shows that the lower deposition temperature result in less intermixing at the Ta-Si interface. 
We fabricate \SI{3}{\micro\meter} gap coplanar waveguide (CPW) resonators and find that the BEOL-window deposited films exhibit state-of-the-art (SOTA) single-photon power quality factors near 4 million.
Finally, we confirm the film performance by demonstrating \SI{20}{\micro\meter} gap transmon qubits with quality factors of 14 million.
We therefore promote Kr as a superior process gas for scalable deposition of Ta films for superconducting qubit applications.

\section{\label{sec:materials} Materials Characterization}

Our magnetron sputtering approach compares Ar and Kr process gases under comparable conditions of gas pressure, 2 mTorr, deposition rate near \SI{0.5}{\nano\meter\per\second}, and substrate temperature (reported from tool thermocouple).
We use \SI{100}{\milli\meter} intrinsic float zone Si(100) wafers and remove the native silicon oxide with a 1 min buffered oxide etchant (BOE) bath prior to loading for deposition.
Details of the thin film deposition are given in Appendix~\ref{app:fab}.

We begin with electronic transport characterization of the films.
Body-centered cubic (BCC) and tetragonal tantalum (\textalpha-Ta and \textbeta-Ta, respectively) have distinct transport signatures that allow quick verification: \textbeta-Ta is more resistive by an order of magnitude at room temperature, has hardly any temperature-dependence in the resistivity, and has a superconducting critical temperature ($T_c$) around \SI{0.5}{\kelvin}. 
In contrast, \textalpha-Ta exhibits the classic temperature dependence of a good metal and has a much higher critical temperature $T_c\approx\SI{4.2}{\kelvin}$ (see Fig.~\ref{fig:tc}).

In Fig.~\ref{fig:intro}c, we show the residual resistivity ratio (RRR) of \SI{100}{\nano\meter} thick films as a function of deposition temperature and sputter gas. 
RRR is the ratio of resistivity at \SI{5}{\kelvin} to that at \SI{305}{\kelvin}; this quantity correlates with the electronic transport mean free path of the material.
Two clear trends are evident. 
First, films deposited with Kr nucleate \textalpha-Ta at a minimum temperature of \SI{200}{\celsius}, whereas Ar requires at least \SI{350}{\celsius}. 
We remark that both threshold temperatures for achieving \textalpha-Ta is not precise, as sometimes the deposition produces \textbeta-Ta.
This may be due to a sensitivity to surface or chamber conditions, which highlights the benefit of a large margin in the process window. 
Second, the Kr-deposited films exhibit substantially higher RRR. 
The room-temperature conductivity is nearly phonon-limited for \textalpha-Ta films made from both sputter gases (Fig.~\ref{fig:transport}), so this shows a much higher transport mean free path in Kr-deposited films at cryogenic temperatures.

To better understand the transport properties, we measure films of varying thickness. 
Combining resistivity and the Hall carrier density, we can estimate the mean free path, $\ell_{\rm 5K}$, (Fig.~\ref{fig:intro}d). 
Films deposited with Kr show consistently higher mean free path which are film thickness-limited, unlike the films deposited with Ar (App.~\ref{app:materials}). 
We additionally measure the temperature-dependence of the critical magnetic field of the superconducting phase, which allows us to estimate the Ginzburg-Landau coherence length $\xi_\mathrm{GL}$. 
We find that the Kr-deposited films show mean free paths consistently larger than $\xi_\mathrm{GL}$, suggestive of films trending towards the clean limit. 
The lower mean free path in Ar-deposited films is consistent with the detection of Ar impurities in the films with secondary ion mass spectrometry (SIMS) (App.~\ref{app:materials}).
Overall, this data demonstrates the superior electronic transport properties of \textalpha-Ta films grown with Kr. 
In Appendix~\ref{app:materials} we show supporting and additional electronic transport data.

\begin{figure*}[t]
\centering
\includegraphics[scale=1]{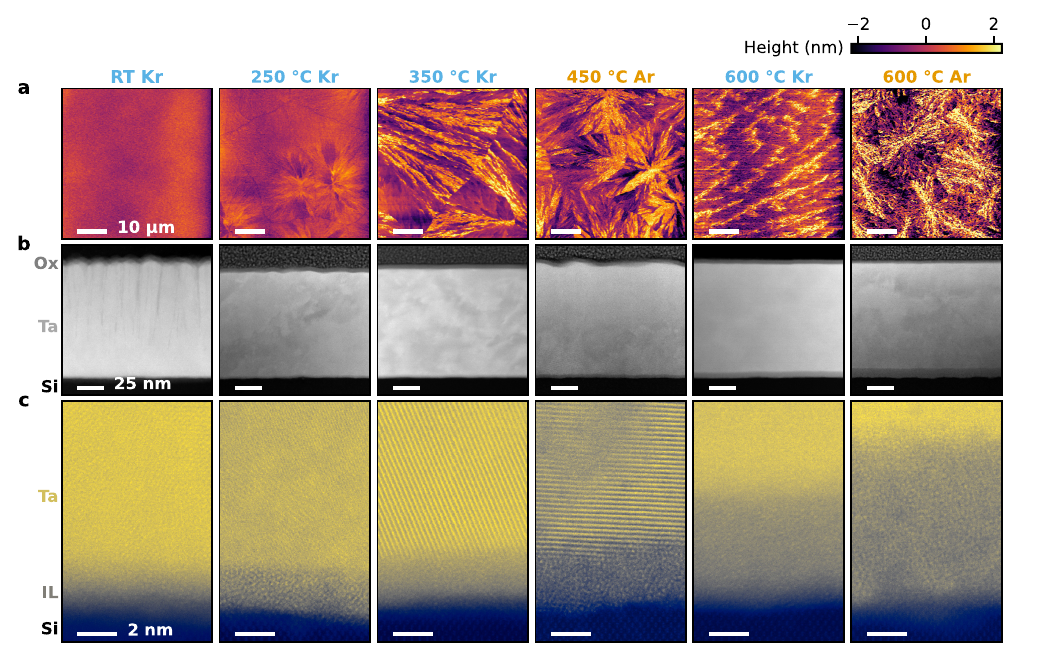}
\caption{\label{fig:structure} Atomic force microscopy (AFM, row a) and scanning transmission electron microscopy (STEM, rows b and c) of Ta films deposited with Kr and Ar between room temperature (RT) to \SI{600}{\celsius}.
AFM and STEM of tantalum films grown with Ar at RT are visually indistinguishable from those grown with Kr at RT.
a) AFM scans with equal-range height colormaps that are zeroed at the mean height of each image.
b) High-angle annular dark-field-STEM (HAADF-STEM) images of the films at low magnification and c) zoomed into the metal-substrate interface, showing the amorphous interlayer between Si and the Ta film. 
A false-color map is applied to improve the perceptibility of the silicon lattice, which is weakly visible in the bottom layer.
}
\end{figure*}

We now turn to the structural characterization of the Ta films, focusing on \SI{100}{\nano\meter} thick films. 
We confirm the crystal phase by X-ray diffraction (XRD) for Ta films deposited at several temperatures and with both sputter gases (App.~\ref{app:materials}). 
Figure~\ref{fig:structure} shows the atomic force microscopy (AFM) and scanning transmission electron microscopy (STEM) data for six key configurations:
Kr-based depositions at room-temperature (RT), \SI{250}{\celsius}, \SI{350}{\celsius}, and \SI{600}{\celsius}, as well as Ar-based depositions at \SI{450}{\celsius} and \SI{600}{\celsius}.
AFM data (Fig.~\ref{fig:structure}a) shows that the Kr-based depositions have a distinct shift in structure between \SI{250}{\celsius} and \SI{350}{\celsius}, transitioning from structures we refer to as `cell'-like to `flower'-like.
Additional detail on this comparison is shown in Fig.~\ref{fig:afmranges}.
These observations are supported by electron back scatter diffraction (EBSD) maps (see App.~\ref{app:materials}), which demonstrate that the typical crystal alignment in the z-axis changes between \SI{200}{\celsius} and \SI{350}{\celsius}.
The change in typical crystal alignments and grain structure indicates a transition of growth mode between the two temperatures.
The EBSD maps reflect the features shown in AFM.
The flower-like features continues up to \SI{600}{\celsius}.

The STEM data (Fig.~\ref{fig:structure}b-c) also show clear effects due to growth at different temperatures. 
Focusing first on the low-magnification images in Fig.~\ref{fig:structure}b, room-temperature growth shows small columnar grains with high-angle boundaries at a pitch of approximately \SI{10}{\nano\meter} scale. 
STEM also reveals substantial surface roughness at short length scales correlated with this columnar growth that is not detected in AFM, and likely too small to be resolved there. 
Oxide columns that penetrate deep into the films are observed only in room temperature samples. 
As growth temperature increases, the spatial density of high-angle grain boundaries lowers significantly, and the short-distance surface roughness progressively improves.

We then examined the Ta/Si interface (Fig.~\ref{fig:structure}c). 
In microwave circuits, this interface has a high electric field participation~\cite{wang_surface_2015, calusine_analysis_2018}, so microwave device performance is highly sensitive to this material region. 
We see clear signs of interlayer formation, and the thickness of this amorphous interlayer grows progressively as substrate temperature increases. 
To clarify the nature of this interlayer, we used electron energy loss spectroscopy (EELS). 
EELS confirms the presence of both Ta and Si in the interlayer, and the changed Si EELS fine structure shows bonding of the Si to Ta, with the absence of a detectable silicate or oxide there.
The interlayer width increases with increasing growth temperature (see Fig.~\ref{fig:eels}). 
In most scans, this interlayer has a smooth chemical gradient, while the high-temperature samples suggest the presence of an explicit silicide.

\begin{figure*}
\includegraphics[width=0.9\textwidth]{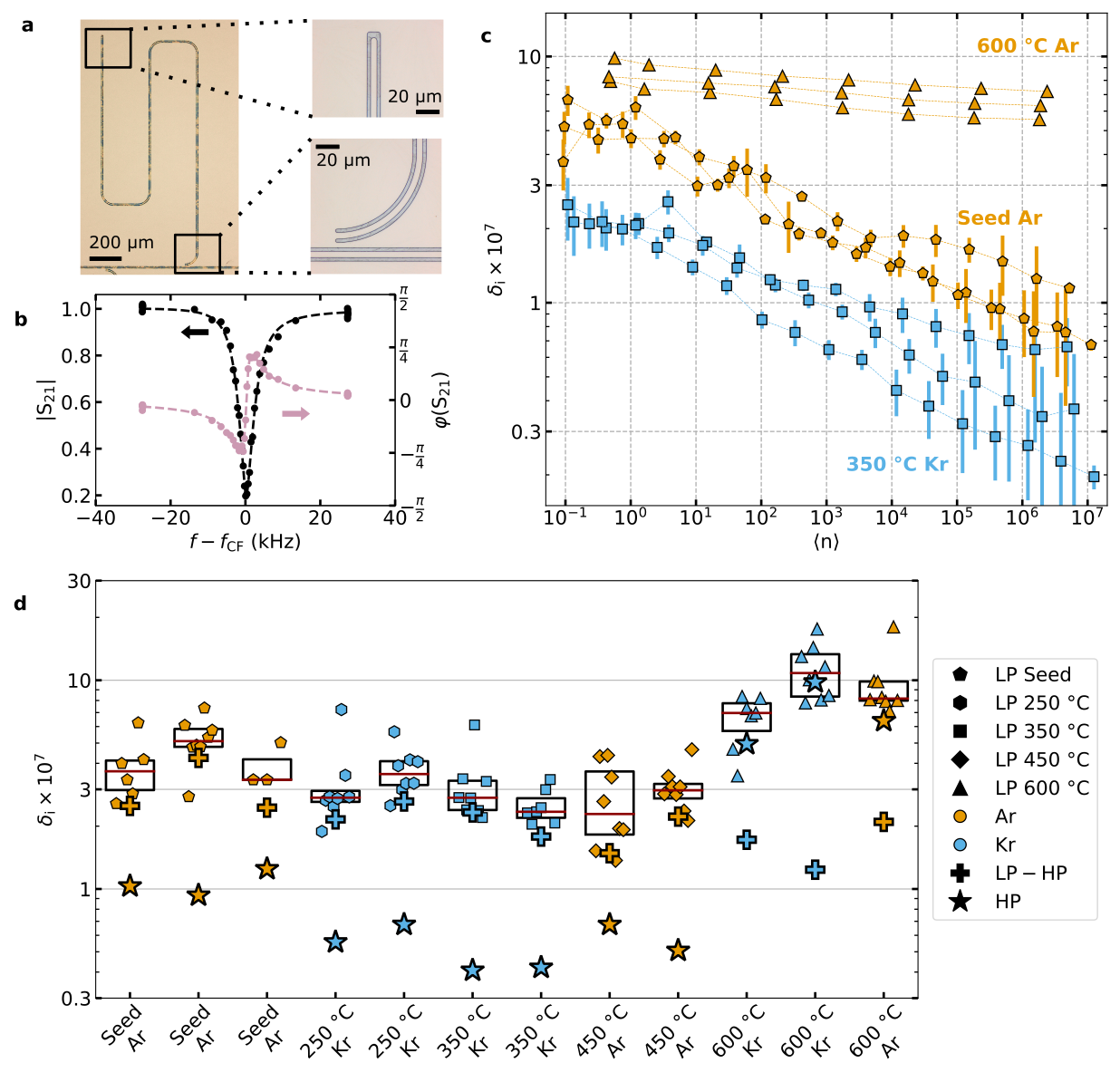}
\caption{
(a) Optical microscope image of a \SI{100}{\nano\meter} thick Ta CPW resonator deposited with Kr at \SI{350}{\celsius}.
The inconsistency of the color in the trenched region reflects the inhomogeneous depth due to grain orientation-dependent etch rates (see also App.~\ref{sec:roughness}). 
(b) Resonator quality factor measurement at single photon powers for \SI{100}{\nano\meter} thick Ta CPW resonator deposited with Kr at \SI{350}{\celsius}.
Black indicates magnitude, while pink is the phase.
(c) Power dependence of losses for three different types of resonators, from top to bottom: \SI{600}{\celsius} Ar, RT Nb-seeded Ar, and \SI{350}{\celsius} Kr.
(d) Resonator losses for samples with post-BOE treatment. 
Each box represents $\delta_\mathrm{LP}$ for one chip: the red lines marks the median LP loss, the black box marks the 25\% and 75\% quartiles.
Stars indicate median HP losses.
Plus signs indicate median difference between LP and HP losses.
\label{fig:resonator}
}
\end{figure*}

\section{\label{sec:microwave}Cryogenic Microwave and Qubit Performance}

We now turn to assessing the performance of microwave devices fabricated from these films. 
We follow closely the fabrication procedures described in our previous work~\cite{olszewski_low-loss_2025} for niobium thin film resonators (a detailed description is given in App.~\ref{app:fab}).
We make coplanar-waveguide (CPW) resonators that derive from the open-source Boulder Cryogenic Quantum Testbed design~\cite{kopas_simple_2022} (Fig.~\ref{fig:resonator}a): a hanger configuration of eight resonators with a \SI{3}{\micro\meter} wide gap between the \SI{6}{\micro\meter} wide center trace and the ground plane. 
This spatially compact structure ensures high electric field participation in the surfaces and interfaces, allowing the device to serve as a sensitive probe of microwave losses in those regions.  

These resonators are measured in a dilution refrigerator (wiring diagram in Fig.~\ref{fig:wiring-diagram}) with standard microwave methods.
A typical single-photon power resonator line shape is shown in Fig.~\ref{fig:resonator}b. 
The $Q_{c}$ (coupled), $Q_{i}$ (internal), and $Q$ (loaded) quality factors were fitted using the diameter correction method (DCM)~\cite{khalil_analysis_2012,mcrae_materials_2020} to
\begin{equation}
    \label{eq:S21-fit}
    S_{21}(f) = 1-\frac{Q/\hat{Q_{c}}}{1+2iQ\frac{f-f_{0}}{f_{0}}}~,
\end{equation}
where $f$ is the probe frequency, $f_{0}$ is the resonance frequency, and $\hat{Q_{c}}=Q_{c}e^{-i\phi}$ is the complex coupled quality factor.
From this, we extract the resonator center frequency, coupling rate, and internal loss $\delta_i=1/Q_{i}$. 
By measuring as a function of input power, which can be converted to an average number of photons in the resonator, we obtain the power-dependence of the internal loss, as shown in Fig.~\ref{fig:resonator}c for several representative resonators. 
The loss decreases with increasing power, as is typical. 

To compare resonators we extract the loss at average energies below one photon in the resonator (low power, LP) and at 100,000 photons (high power, HP).
The LP losses serve as a benchmark in the regime of energy relevant for qubit performance.
The HP losses describe power-independent losses, the known physical mechanisms for which include charged defect-mediated phonon emission~\cite{turiansky_dielectric_2024}, interface piezoelectric losses~\cite{zhou_observation_2024}, and quasiparticles~\cite{crowley_disentangling_2023}.
For the lowest HP losses, packaging issues may also enter~\cite{rieger_fano_2023}.
The power-dependent losses (the gap between LP and HP) are typically ascribed to two-level systems~\cite{de_leon_materials_2021}, although here we do not observe the 'S-curve' predicted by classic models~\cite{mcrae_materials_2020,altoe_localization_2022,verjauw_investigation_2021}.

Figure~\ref{fig:resonator}d shows the comparison of a series of resonators from different films, all of which were treated with a BOE `cleanup' step, which typically improves performance, around three hours before loading into the dilution refrigerator. 
We find that the resonators deposited directly on Si at lower temperatures (\SI{250}{\celsius} to \SI{450}{\celsius}) exhibit best-in-class performance, comparable or better than in~\cite{marcaud_low-loss_2025}. 
The resonators made from films deposited with Kr at \SI{350}{\celsius}, which we use later for qubit devices, exhibited median resonator quality factors of 4.1 million at low power and 25 million at high power. 

The resonators made from films deposited at \SI{600}{\celsius} are limited by HP losses. 
Remarkably, for these films, we sometimes find the BOE treatment \textit{increases} the HP losses by an order of magnitude (Fig.~\ref{fig:noBOE}. 
This is potentially related to hydrogen incorporation, which was recently observed to impact Ta films subjected to heavy doses of hydrogen fluoride~\cite{lozano_reversing_2025}.
Moreover, several samples fabricated from these films resulted in extremely high losses ($\delta_\mathrm{HP}\gtrsim10^{-5}$). 
At this time, it remains unclear why the \SI{600}{\celsius} films exhibit inconsistent or BOE-sensitive microwave losses.
The only clear differences observed in the materials characterization are the thickness of the intermixed layer at the metal-substrate interface and the grain boundary density.
This provides a correlation for future investigation. 

Finally, we also compare with RT-deposited Ta films that were seeded by a \SI{5}{\nano\meter} layer of Nb to stabilize \textalpha-Ta.
Comparing to the intermediate-temperature, direct-deposited films, the seeded-film resonators exhibit marginally higher LP losses (consistent with a recent report~\cite{marcaud_low-loss_2025}) and approximately double the HP loss.
In this case, however, the BOE treatment and rapid transfer to vacuum conditions are critical. 
After aging for a few weeks, the LP losses increase by a factor of three (Fig.~\ref{fig:aging}) and the HP losses increase by almost a factor of two.
In contrast, aging the direct-deposited films only results in increased LP loss, and the direct-deposited films remain overall better-performing.

The surface oxide of Ta is known to be relatively stable over time and to BOE treatment~\cite{crowley_disentangling_2023}.
In contrast, Nb and Si surface oxides are removed by BOE treatment and then regrow over time~\cite{verjauw_investigation_2021,altoe_localization_2022}.
We, therefore, propose that the regrowth of the Si surface oxide contributes to aging-induced power-dependent losses, while the aging-induced power-dependent losses in the seeded films may be associated to the film corners with exposed Nb.
This interpretation is supported by data suggesting similar aging effects in all-Nb resonators~\cite{verjauw_investigation_2021}

\begin{figure*}
\includegraphics[width=\textwidth]{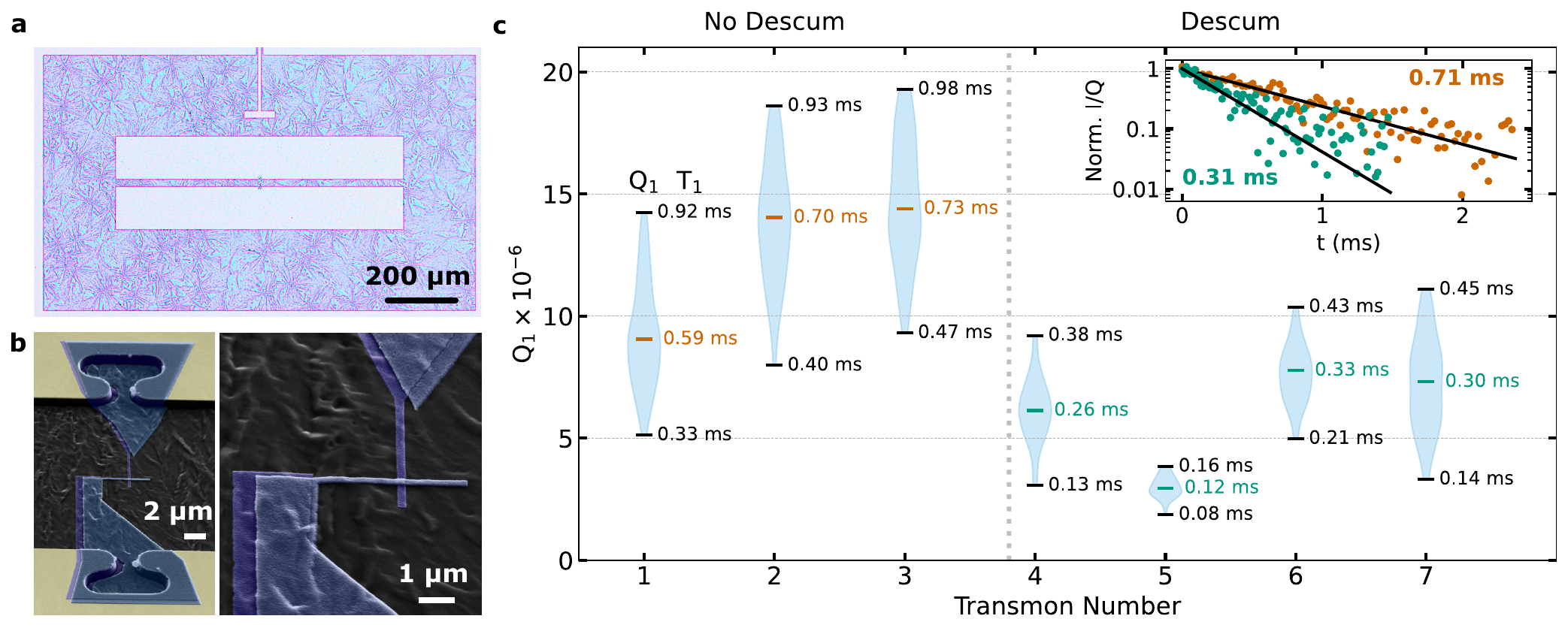}
\caption{
\label{fig:transmon}
(a) Optical microscope image of the transmon device with a \SI{20}{\micro\meter} gap between the capacitor pads.
The inconsistency of the color in the exposed Si reflects the inhomogeneous depth due to grain orientation-dependent etch rates (see also App.~\ref{sec:roughness}).
(b) False-colored scanning electron microscopy images of the region of the Josephson junction. The first and second Al electrodes of the junction are show in purple and blue, respectively. The tantalum film is shown in yellow.
(c) Violin plots of transmon quality factor for seven qubits measured; the distributions come from measuring $T_1$ repeatedly over about twelve hours.
Values top to bottom represent maximum, median, and minimum lifetimes measured.
Transmons 1-3 were fabricated without an oxygen descum, while transmons 4-7 had an additional oxygen descum.
Insert has two example plots of lifetimes near the median value for both sets of devices, Norm. I/Q is the normalized I/Q signal and t is the measurement time.
}
\end{figure*}

Finally, we close the loop by fabricating transmon devices utilizing the Ta films deposited at \SI{350}{\celsius} with Kr sputter gas.
The transmon capacitor pads are rectangular and separated by a \SI{20}{\micro\meter} gap (Fig.~\ref{fig:transmon}a).
The fabrication procedure follows the same recipe as the resonators, except that we etch deeper for these devices: approximately \SI{500}{\nano\meter} rather than \SI{100}{\nano\meter}. 
This was motivated by studies showing that deeper trenches mitigate surface participation ratio~\cite{wang_surface_2015, calusine_analysis_2018}. 
We did not find that this trenching process improves performance for the CPW resonators (Fig.~\ref{fig:deep}), but this may still mitigate surface losses for full qubit fabrication because BOE-based cleaning is not possible after depositing the Josephson junctions. 
The Josephson junctions are fabricated with a standard shadow-evaporation method in the Manhattan style with an Al/AlO\textsubscript{x}/Al superconductor-insulator-superconductor structure (Fig.~\ref{fig:transmon}b)~\cite{bland_millisecond_2025,place_new_2021,tuokkola_methods_2025,kono_mechanically_2024}. 
We report two datasets from devices with Josephson junctions, fabricated with and without a descum process prior to loading into the evaporation chamber, motivated by previous work describing the utility of descum processes for improving uniformity or other performance metrics~\cite{kriekebaum_superconducting_2020,gingras_improving_2025}.
Other than this, the chips were produced in parallel with the same nanofabrication processes and from the same wafer.

We characterize the transmons by measuring their relaxation and dephasing times via standard time-domain methods.
In Fig.~\ref{fig:transmon}, we show a typical relaxation decay curve for a qubit with the descum and and a qubit without the descum, indicating $T_1\approx \SI{0.71}{\milli\second}$ for the no-descum device and $T_1\approx \SI{0.31}{\milli\second}$ for the with-descum device. 
For each transmon, we collected data over approximately twelve hours to capture the time-domain fluctuations, which are condensed into distributions as shown in Fig.~\ref{fig:transmon}d. 
We show both $T_1$ and quality factor $Q_1=2\pi f_{\rm qubit} T_{1}$ (which normalizes lifetime to the qubit frequency).
The qubits on the no-descum chips exhibit median quality factors as high as 14.4 million, within 6\% of the best-reported Ta-on-Si transmon qubits (15.2 million)~\cite{bland_millisecond_2025}. 
For this $Q_1$, $T_1$ is below the \SI{1}{\milli\second} threshold due to the higher qubit frequency. 
Notably, the capacitors of our devices are 3.5 times closer together, which makes the qubits substantially more sensitive to surface and interface losses. 
These results demonstrate that the Kr-deposited films are compatible with the full qubit fabrication workflow and also validate our overall nanofabrication process.

The chips exposed to the descum step exhibited substantially lower performance, with three of four qubits exhibiting $Q_1$ between 6 and 7 million. 
The fractional distribution of the relaxation times is comparable for all qubits, ranging from -41\% to +22\%.
Remarkably, the performance gap for dephasing is inverted: the qubits chip exposed to the descum process exhibited over twice the dephasing times than the qubits on the other chip (Hahn echo times averaging \SI{0.22}{\milli\second} vs \SI{0.1}{\milli\second}, see Fig.~\ref{fig:transmon_t2}). 
This discrepancy highlights that the details of the JJ fabrication remain important and will be investigated in more detail in future work.

\section{Discussion}

These results open the door to qubit development research following a similar thread of combining high performance materials with scalable device production methods. 
We point out a few immediately identifiable directions. 
First, exploring additional sputter gases such as neon and xenon~\cite{benvenuti_study_1999,matson_effect_2000,lee_high-rate_2002} will help in further illuminating interplay of the sputter gas with the nucleation dynamics, impurities, and ultimate microwave performance for Ta-based films.
We note that Ar-deposited films show detectable levels of Ar impurities while Kr impurities are not observed in any films.
One possibility is that Ar and Kr have different measurement yields in SIMS, a plausible explanation we cannot exclude at this time.
However, the differences in resistivity and mean free path between Ar- and Kr-deposited films indicate that Kr-based films have a lower density of transport scattering sites.
It is also predicted that the momentum transfer during sputtering between Kr and Ta is more efficient as compared to Ar and Ta~\cite{tong_advanced_2014}.
The effect of momentum transfer has been extensively studied for various transition metals, such as V and W~\cite{petrov_comparison_1993,paturaud_influence_1996}, suggesting that Kr is less likely to incorporate to the Ta film.
We hypothesize that this lower propensity for noble gas impurities helps to unlock the benefits of high kinetic energy target atoms produced by sputtering to nucleate the thermodynamically stable BCC phase.
The previous observations of benefits for other materials~\cite{benvenuti_study_1999} also suggests that the ongoing efforts to identify novel materials and alloys for superconducting qubits~\cite{ganjam_improving_2023,choi_low_2025,yang_superconducting_2025} will benefit from exploration of alternative sputter gases like Kr.

We also highlight the potential for this approach to expand the range of substrate and thin film surfaces on which BCC Ta can be deposited, with particular application towards Josephson junctions. 
It is known that the high nucleation temperature for \textalpha-Ta films with Ar-based sputtering applies not just to Si surfaces but to many others, including sapphire~\cite{place_new_2021,bahrami_vortex_2025} and various transition metals and their oxides~\cite{feinstein_factors_1973,sato_nucleation_1982,urade_microwave_2024,zikiy_investigation_2025}.
Kr-based sputtering may lower the barrier for these and other surfaces, too. 
For example, trilayer heterostructures are a common target for making Josephson junctions with only subtractive shape-defining techniques, as preferred by  industry~\cite{anferov_improved_2023,choi_low_2025,sethi_native-oxide-passivated_2025,bhatia_ta-based_2025}.
Using Ta electrodes would open the ultimate device to more aggressive cleaning than is possible with Al-based junctions~\cite{urade_microwave_2024,zikiy_investigation_2025,crowley_disentangling_2023,lozano_low-loss_2024} while also taking advantage of the self-limiting and relatively low-loss surface oxide of Ta in the immediate vicinity of the junction.
Heavy gas-based sputtering deposition may lower the temperature threshold for depositing \textalpha-Ta onto a variety of different prospective tunnel barrier materials, such as AlO\textsubscript{x}~\cite{anferov_improved_2023},  ZrO\textsubscript{x}~\cite{choi_low_2025}, and Ta\textsubscript{2}O\textsubscript{5}~\cite{potluri_fabrication_2025}, thereby opening the door to more materials exploration and a more direct path to scaling processes to semiconductor facilities. 

In conclusion, we showed that switching from Ar-based to Kr-based Ta growths substantially enhances the material and transport quality of our thin films.
This improvement, coupled with lower deposition temperatures, coincides with an increase in grain size, reduction of Ta/Si intermixing, decrease in argon impurities, and enhancement of conductivity.
These benefits occur simultaneously with a lower temperature threshold for creating \textalpha-Ta films, providing a clear pathway for adapting this technique in large-scale fabrication, particularly for foundry-scale \SI{300}{\milli\meter} fabrication that accomplish metal-based fabrication below \SI{400}{\celsius}.
These characteristics favor adoption of Kr-based Ta depositions for high-quality superconducting device applications, such as superconducting resonators and qubits.

\paragraph*{Author Contributions}
M.W.O. conceptualized the experiment with guidance from V.F..
M.W.O. adapted the deposition chamber and developed the thin film deposition methods with help from L.K..
L.K. accomplished the AFM measurements with guidance from V.F. and D.A.M. and help from M.W.O. and S.R..
L.K. accomplished the XRD, and EBSD measurements with guidance from V.F. and D.A.M..
D.T. accomplished the TEM measurements with guidance from D.A.M..
A.B.B. accomplished the SIMS measurements.
M.W.O. accomplished the transport measurements with help from L.K. and S.R..
M.W.O. fabricated the resonators with help from L.K..
M.W.O. conducted the resonator measurements with help from H.L..
G.D.G., X.D., S.R., and L.Z. developed the JJ fabrication with help from M.W.O..
M.W.O. and S.R. fabricated the transmons with help from X.D. and L.K..
S.R. and M.W.O. conducted the transmon measurements with help from S.~Roy.
S.R. set up the measurement cryostat with help from L.K. and M.W.O..
V.F. supervised the project.
M.W.O. and V.F. led the manuscript writing with help from L.K., S.R., and D.T..
All authors reviewed and provided feedback on the manuscript.

\begin{acknowledgments}
We are grateful for help on this work from Kushagra Aggarwal on nanofabrication, and Tathagata Banerjee on XPS.
We thank multiple individuals for helpful contributions: Anthony P. McFadden and Corey Rae H. McRae (NIST-Boulder) for sharing their transmon qubit design, which we adapted for this study; Faranak Bahrami and Nathalie P. de Leon (Princeton) for discussions on \textalpha-Ta and JJ fabrication; Satyavolu Papa Rao (NY Creates) for discussions on scalable nanofabrication; Gregory D Fuchs (Cornell) for discussions throughout the project; Melissa Hines (Cornell) for discussions on proper silicon preparation techniques; Dan Ralph (Cornell) for access to essential equipment; and Lopa Bhatt (Cornell) for help collecting and interpreting electron energy loss spectrometry data.
We also thank Christopher Alpha, Jeremy Clark, Steve Kriske, Tom Pennell, Mark Pfeifer, and Aaron J. Windsor (technical staff at at the Cornell Center for Materials Research and the Cornell Nanoscale Facility) for assistance in materials analysis and standing up fabrication processes.
We acknowledge help from AJA International, Inc.,~technical staff, especially Wendell Sawyer.

This prototype was primarily supported by the Microelectronics Commons Program, a DoW initiative, under award number N00164-23-9-G061.
Funding for shared facilities used in this prototype was provided by the Microelectronics Commons Program, a DoW initiative, under award number N00164-23-9-G061.
This work was performed in part at the Cornell NanoScale Facility, a member of the National Nanotechnology Coordinated Infrastructure (NNCI), which is supported by the National Science Foundation (Grant NNCI-2025233).
This work made use of the Cornell Center for Materials Research shared instrumentation facility.
This work made use of the Meehl cryostat donated by David W. Meehl in memory of his father James R. Meehl and supported by the Cornell College of Engineering.

\end{acknowledgments}

\clearpage

\appendix

\renewcommand{\thefigure}{A\arabic{figure}}
\setcounter{figure}{0}

\section{Device Fabrication Details}~\label{app:fab}
\subsection{Thin Film Synthesis}
Our fabrication is adapted from~\cite{olszewski_low-loss_2025}.
We use commercially available, double side polished, float-zone, high resistivity ($\geq$\SI{10}{\kilo\ohm}), intrinsic, 4-inch silicon wafers from WaferPro.
The day of the deposition, the wafers are cleaned with a two step RCA clean: H$_{2}$O:NH$_{4}$OH:H$_{2}$O$_{2}$ (6:1:1) and H$_{2}$O:HCl:H$_{2}$O$_{2}$ (6:1:1) at \SI{70}{\celsius} for \SI{10}{\min} each.
\SI{3}{\min} before loading into the deposition chamber the native silicon oxide is removed with a \SI{1}{\min} 10:1 BOE treatment, followed by a \SI{30}{\second} rinse in Milli-Q water.
The glassware and equipment used for the BOE processing is periodically cleaned with a two-step RCA clean to minimize contamination.
The load lock is pumped down to below $<10^{-6}$ Torr. The sample is then transferred into the deposition chamber, which has a typical base pressure $<5\times10^{-9}$ Torr.

The Ta base layer depositions were performed in a chamber traditionally used for spintronics research (AJA International) as discussed in our previous work~\cite{olszewski_low-loss_2025}.
All Ta-only depositions were done by co-sputtering with two 2-inch diameter and 1/4-inch thick targets with 4N5 purity from AJA International.
We did not find any meaningful differences in material characteristics of films deposited with one or two guns, as shown in Sec.~\ref{sec:resistivity}.
The throw distance between the targets and the sample is around \SI{15}{\centi\meter}, and the substrate is rotated at 50 RPM.
All depositions are done with 2 mTorr and 70 sccm of research plus grade gas from Airgas.
Ar-based depositions used a \SI{400}{\watt} bias (\SI{876}{\milli\ampere} and \SI{455}{\volt}), and Kr-based depositions were done with a \SI{300}{\watt} bias (\SI{581}{\milli\ampere} and \SI{514}{\volt}), for each gun.
Both settings resulted in a similar deposition rate of approximately \SI{0.54}{\nano\meter\per\second}.
Heated depositions were heated to target temperature and equilibrated for \SI{20}{\min} prior to deposition.
We report the substrate temperature as seen on the internal tool thermocouple.
We verified the substrate temperature with both non-reversible thermometer strips and a bi-metal lazy hand thermometer at a few temperatures.
We find the that the reported temperature is within \SI{-10}{\celsius} to \SI{+60}{\celsius} of the measured temperature using the calibration methods in \SI{200}{\celsius} to \SI{350}{\celsius} temperature range.
The films were \SI{100}{\nano\meter} thick unless otherwise indicated.

For the Nb-seeded films, a \SI{5}{\nano\meter} Nb seed layer was deposited, followed by \SI{100}{\nano\meter} Ta films.
All of the seeded depositions were done at room temperature with a single gun deposition for both niobium and tantalum using argon process gas.

\subsection{Resonator and Base Layer Fabrication}
After deposition, the films were coated with a single layer MICROPOSIT S1813 resist with 3000 RPM spin rate, and baked at \SI{90}{\celsius} for \SI{1}{\min}.
The resonator and base layer features were exposed with a g-line stepper tool, developed in an MIF-based developer, and cleaned with a mild oxygen plasma for \SI{90}{\second} in a reactive ion etching (RIE) tool.
The samples were then etched with a chlorine-based chemistry (30:20:5 sccm for Cl$_{2}$:BCl$_{3}$:Ar at 7 mTorr) in a RIE with an inductively coupled plasma.
The approximate etch rate is \SI{1.5}{\nano\meter\per\second}.
We note that the silicon surface is roughened by the etch, as discussed in detail in Sec.~\ref{sec:roughness}.

After etching, the resist is stripped with a \SI{1}{\hour} \SI{85}{\celsius} treatment in Integrated Micro Materials AZ300T stripper with agitation by a magnetic stir bar.
This is followed by a series of room temperature sonications in isopropanol (\SI{10}{\min}), Milli-Q water (\SI{10}{\min}), Milli-Q water (\SI{1}{\min}), and isopropanol (\SI{1}{\min}).
The samples are then dried with compressed nitrogen.
The glassware used for the resist stripping is periodically cleaned with a two-step RCA clean.

For resonator fabrication, we then coat the wafers with a protective layer of S1813 resist for dicing.
After dicing the samples into \SI{6.5}{\milli\meter} by \SI{6}{\milli\meter} chips, we strip the protective dicing resist within \SI{6}{\hour} of loading into a cryostat. Once again we use a \SI{1}{\hour}, \SI{85}{\celsius} treatment in AZ300T stripper with magnetic stir bar agitation.
This is followed by a series of room temperature sonications in AZ300T (\SI{10}{\min}), isopropanol (\SI{10}{\min}), Milli-Q water (\SI{10}{\min}), Milli-Q water (\SI{1}{\min}), and isopropanol (\SI{1}{\min}).
The samples are then dried with compressed nitrogen.
Samples with a post-fabrication BOE treatment are then dipped in 10:1 BOE for \SI{1}{\min} followed by two \SI{30}{\second} rinses in Milli-Q water and a nitrogen blow dry.
The BOE step is typically  done within \SI{3}{\hour} of loading samples into the cryostat.
The sample is then bonded to our package with aluminum wire bonds.

\subsection{Transmon Fabrication}
After etching the Ta base layer, the photoresist is stripped and the wafer is treated with a \SI{1}{\min} 10:1 BOE bath, followed by two \SI{30}{\sec} water rinses.
The wafer is then spin coated with a bilayer of e-beam lithography (EBL) resist: MMA EL6 and PMMA 495k A8, both at 2100 RPM. Each layer is baked at \SI{145}{\celsius} for \SI{5}{\min}.
The EBL is done with a JEOL 6300 series tool with a dose of \SI{1000}{\micro\coulomb\per\centi\meter\squared}.
After the EBL the wafer is hand cleaved into \SI{19}{\milli\meter} by \SI{18}{\milli\meter} coupons, and then developed with 3:1 IPA:MIBK (isopropanol:methyl isobutyl ketone) for \SI{2}{\min} without agitation, followed by a \SI{30}{\sec} rinse in IPA, and a nitrogen blow dry.
At this stage samples with the additional descum treatment were treated with a \SI{90}{\second} oxygen plasma (\SI{75}{\watt}, 50 sccm, 20 mTorr) in an Oxford 81 RIE tool.
The sample is then loaded into the Angstrom Quantum Series electron beam evaporator equipped with separate oxidation and evaporation chambers.
The typical base pressure is $<2\times10^{-9}$ Torr, and $<2\times10^{-8}$ Torr for the evaporation, and oxidation chambers, respectively.
Aluminum with 5N purity is deposited directly from the copper hearth without using a crucible.
Our deposition follows a typical Manhattan style Josephson junction deposition, as outlined in~\cite{bland_millisecond_2025}.

The sample is first milled with Ar plasma in order to remove tantalum and silicon oxide for making good contact between the tantalum capacitor and the aluminum electrodes of the Josephson junction.
We perform a two step milling process at \SI{45}{\degree} tilt angle, oriented along the axis of each of the following aluminum depositions.
Each step is done by milling twice for \SI{30}{\second} at a beam voltage of \SI{200}{\volt} with a \SI{12}{\second} break in-between.
After ion milling, the first layer of aluminum is deposited for \SI{35}{nm} at \SI{0.2}{\nano\meter\per\second}.
Followed by a \SI{5}{\min} oxidation at 10 Torr of oxygen.
The second deposition is done at \SI{90}{\degree} rotation with respect to the first layer for \SI{70}{nm} at \SI{0.2}{\nano\meter\per\second}.
We perform a final oxidation of \SI{10}{\min} at 10 Torr of oxygen.

The sample is then coated with a layer of S1813 photoresist for protection during dicing, and diced into \SI{6.5}{\milli\meter} by \SI{6}{\milli\meter} chips.
The smaller chips are then lifted-off and stripped in a \SI{1}{\hour} \SI{85}{\celsius} bath of 99.9\%+ pure dimethyl sulfoxide (DMSO).
Followed by a second \SI{1}{\hour} \SI{85}{\celsius} bath of DMSO.
Next the samples are rinsed in IPA, Milli-Q water, Milli-Q water, and IPA.
Finally the samples are blown dry with nitrogen compressed gas.
The samples are then packaged and loaded into the cryostat for measurements.

\section{Additional Materials Characterization}\label{app:materials}

\subsection{Resistivity}
\label{sec:resistivity}
\begin{figure}
\centering
\includegraphics[width=0.8\linewidth]{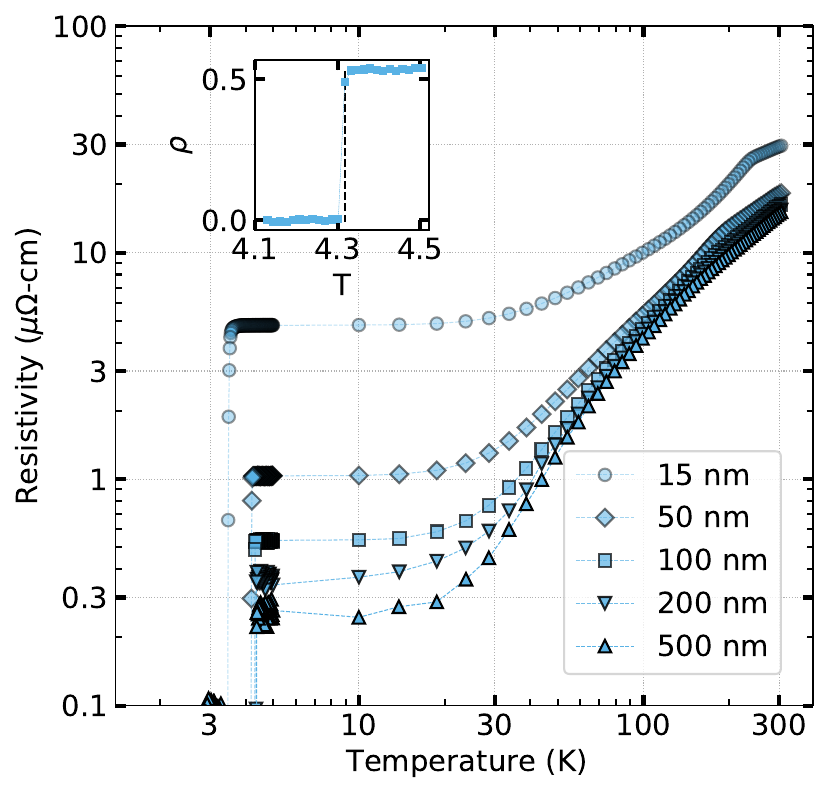}
\caption{\label{fig:tc}
Measurements of resistivity and superconducting transition temperature $\rm T_{\rm c}$ for Ta films deposited at \SI{350}{\celsius} with krypton with various thicknesses.
The insert shows a zoomed in portion of the plot near $\rm T_{\rm c}$ for a \SI{100}{\nano\meter} film.
The $\rm T_{\rm c}$ for this film was \SI{4.32}{\kelvin}.
}
\end{figure}

\begin{figure}
\includegraphics[width=\linewidth]{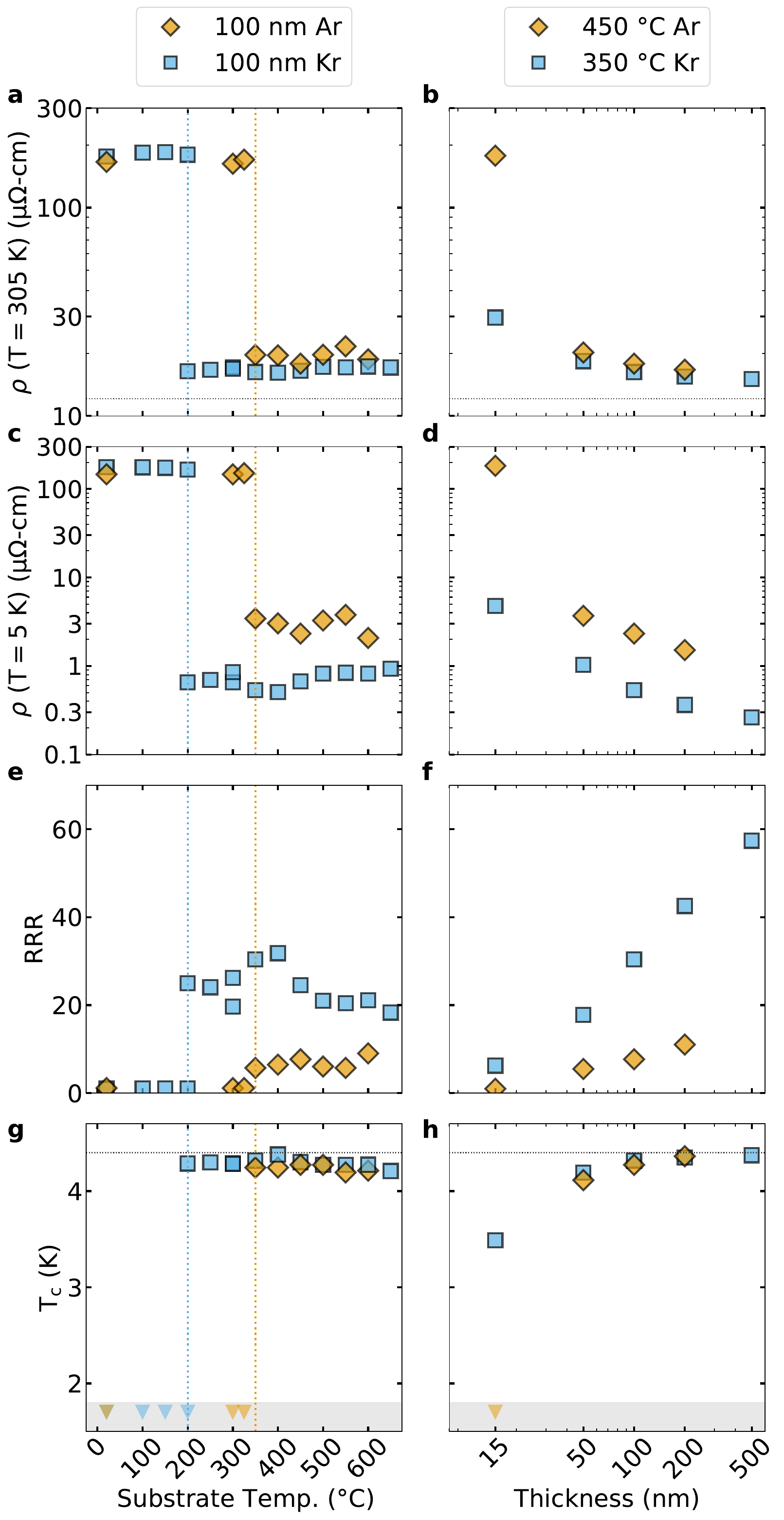}
\caption{\label{fig:transport}
Transport measurements of tantalum films deposited with krypton (blue squares) and argon (orange diamonds) process gasses.
Left column has \SI{100}{\nano\meter} samples deposited at different substrate temperatures.
Right column has samples deposited at \SI{350}{\celsius} for krypton and \SI{450}{\celsius} for argon with different thicknesses.
Panels (a) and (b) plot resistivity measured at \SI{305}{\kelvin}, the dashed line represents the value for bulk resistivity of \SI{12.08}{\mu\Omega\centi\meter}~\cite{sauerzopf_anisotropy_1987}.
Panels (c) and (d) plot resistivity at \SI{5}{\kelvin} for which the bulk value is \SI{0.018}{\mu\Omega\centi\meter}~\cite{sauerzopf_anisotropy_1987}.
Panels (e) and (f) show residual resistivity ratio (RRR), the ratio of the two resistivities.
Panels (g) and (h) are the superconducting critical temperature, where we were unable to measure the transition temperature below \SI{1.8}{\kelvin}.
We denote the samples were we did not measure superconductivity with a triangle at \SI{1.8}{\kelvin}.
}
\end{figure}

Resistivity measurements are summarized in the main text in Fig.~\ref{fig:intro}.
The measurements were done on unpatterned samples which were hand cleaved to a square of about \SI{5}{\milli\meter} by \SI{5}{\milli\meter}.
The measurements were done with a van der Pauw geometry, from which we extract the value for sheet resistance.
Examples of resistivity versus temperature curves are shown in Fig.~\ref{fig:tc}.

\begin{figure}
\centering
\includegraphics[width=\linewidth]{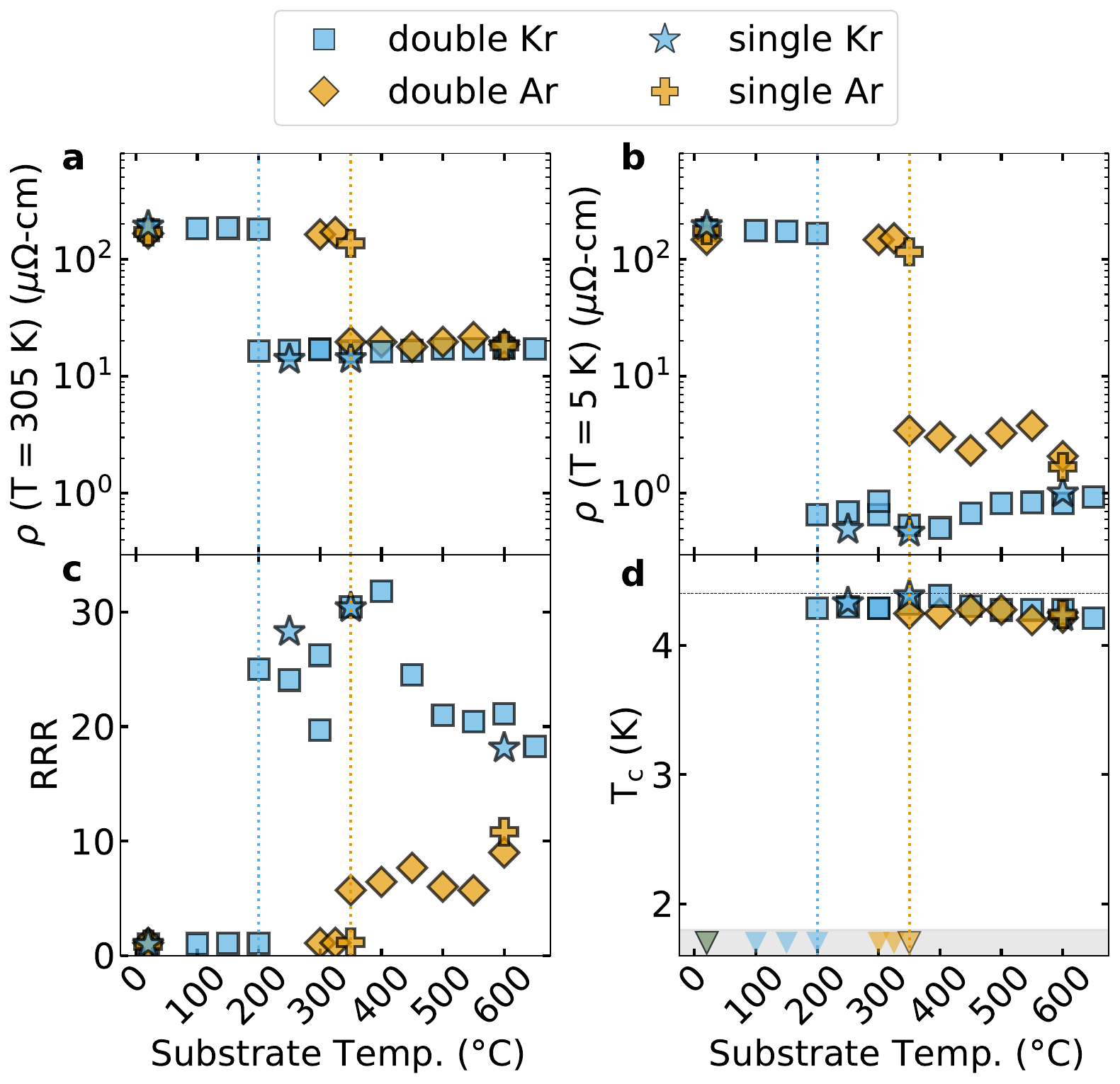}
\caption{\label{fig:res_single}
Resistivity measurements, similar as Fig.~\ref{fig:transport}, with the addition of depositions done with one tantalum gun (single) and depositions done with co-sputtering of two tantalum guns (double).
(a) Measurements of resistivity at \SI{305}{\kelvin} ($\rho$ (305K)).
(b) Measurements of resistivity at \SI{5}{\kelvin} ($\rho$ (5K)).
(c) RRR, and (d) T$_{\rm c}$.
}
\end{figure}

As discussed in the fabrication recipe, most of our depositions were done by co-sputtering with two tantalum targets.
Based on resistivity measurements shown in Fig.~\ref{fig:res_single}, we did not see a meaningful difference between samples deposited with a single gun versus a double gun approach, with otherwise equivalent deposition parameters. 
For the bulk of the study we focused on depositions done with the double gun approach.

\subsection{Critical Magnetic Field}
\begin{figure}
\centering
\includegraphics[scale=1]{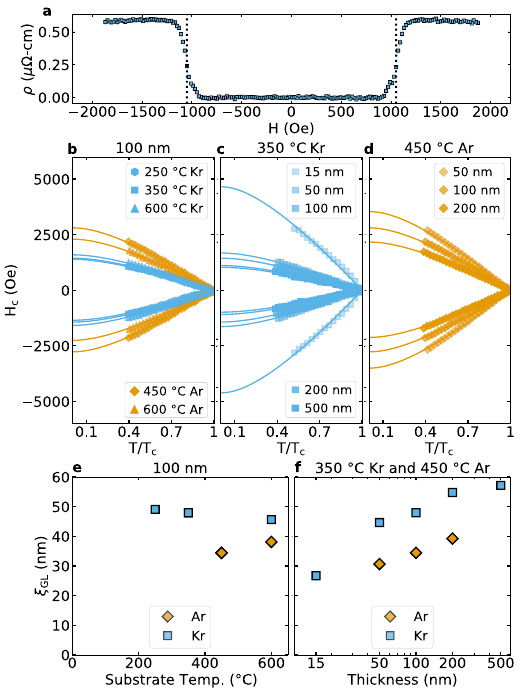}
\caption{\label{fig:critical}
(a) Sample measurement of critical field ($\rm H_{\rm c}$) for a \SI{350}{\celsius} \SI{100}{\nano\meter} krypton film at \SI{1.7}{\kelvin}.
The dotted vertical lines highlight the extracted positive and negative critical magnetic fields.
(b) $\rm H_{\rm c}$ versus temperature for various \SI{100}{\nano\meter} films, fitting is done using equation~\ref{eq:critical}.
(c) $\rm H_{\rm c}$ versus temperature for \SI{350}{\celsius} krypton samples for different thicknesses.
(d) $\rm H_{\rm c}$ versus temperature for \SI{450}{\celsius} argon samples for different thicknesses.
(e) Measured Ginzburg-Landau coherence length ($\xi_{\rm GL}$) for \SI{100}{\nano\meter} samples deposited with different temperatures and gasses. We calculate the value using equation~\ref{eq:gl}.
(f) $\xi_{\rm GL}$ for different thicknesses of \SI{350}{\celsius} krypton and \SI{450}{\celsius} argon films.
}
\end{figure}

Similarly to resistivity measurements, critical magnetic field measurements are done on unpatterned, hand-cleaved samples.
Resistivity as a function of the out-of-plane magnetic field is measured using the van der Pauw technique.
The critical field $\rm H_{\rm c}$ is estimated using the inflection points of $\rho(H)$, as shown in Fig.~\ref{fig:critical}(a).
From $\rm H_{\rm c}$ versus temperature data, shown in Fig.~\ref{fig:critical}(b), we extract the $\rm H_{\rm c}$ at \SI{0}{\kelvin} by numerically fitting to the formula
\begin{equation}
    \label{eq:critical}
    H_{c}(T) = H_{c}(T=0) \frac{\mathrm{Un}(T)}{\mathrm{Un}(T=0)}~,
\end{equation}
where $\mathrm{Un}(T)$ is the universal function defined by de Gennes~\cite{de_gennes_superconductivity_1999, harper_mixed_1968}.
This analytical form of $H_c(T)$ was originally found for superconducting films in the extreme dirty limit.
It is, however, known that the form of the curve is not sensitive to the purity of the superconductor~\cite{helfand_temperature_1966}. 
Indeed, we observe good fits using this function for all of our films, and therefore Eq.~\ref{eq:critical} allows us to extrapolate for values of $H_c(T=0)$.
From the fit, we extract the Ginzburg-Landau coherence length $\xi_{\rm GL}$ using
\begin{equation}
    \label{eq:gl}
    \rm H_{c}(0) = \frac{\phi_{0}}{2\pi\xi_{\rm GL}^{2}}~,
\end{equation}
where $\phi_{0}= h/(2e) = 2.07\times10^{-15}~\si{\weber}$ is the magnetic flux quantum. 
All Kr-deposited films thicker than \SI{15}{\nano\meter} are found to have longer coherence lengths than Ar-deposited films.

\subsection{Mean Free Path}
\begin{figure}
\centering
\includegraphics[width=\linewidth]{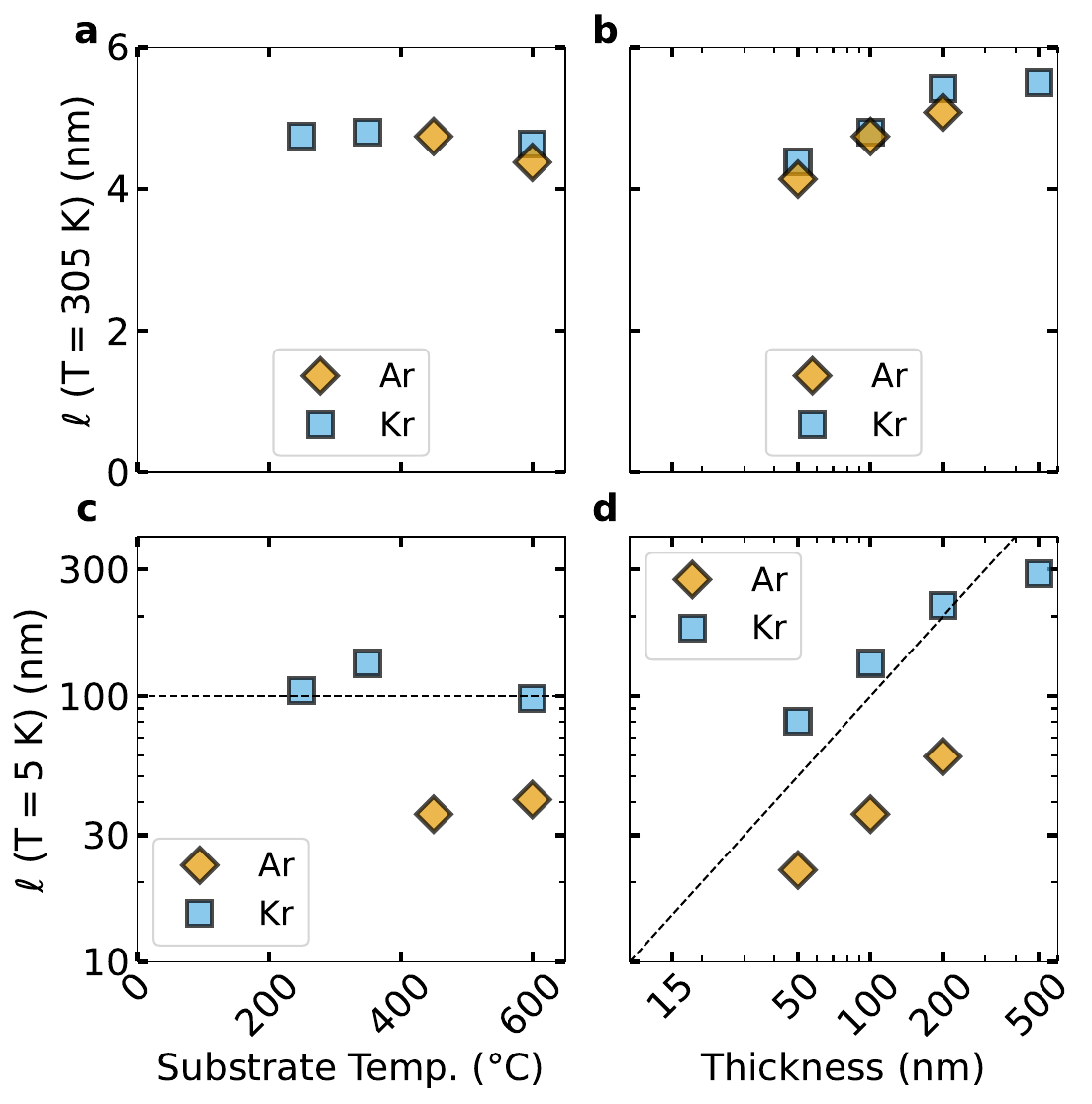}
\caption{\label{fig:mfp}
(a) and (b) Mean free path measured at room temperature ($\ell_{\rm 305K}$) plotted against substrate temperature during deposition, and thickness respectively. (c) and (d) Mean free path measured at \SI{5}{\kelvin} ($\ell_{\rm 5K}$).
The black dashed line in panels (c) and (d) is plotted along $\rm d=\ell$ to indicate roughly the condition where the mean free path is surface-scattering limited.
}
\end{figure}

Using a van der Pauw geometry and unpatterned films, we measure the Hall resistivity versus perpendicular magnetic field to extract the Hall coefficient from the slope: $\rm R_{H} = \rho_{\rm xy}/H$.
From this, we obtain the Hall carrier density, $n_{\rm H}=1/e\rm R_{\rm H}$, which allows us to estimate the product:
\begin{equation}
    \label{eq:mfp}
    \rho\ell=\frac{mv_{\rm F}}{e^{2}n_{\rm H}},
\end{equation}
where $\rho$ is the normal state longitudinal resistivity, $\ell$ is the electron mean free path, $m$ is the electron effective mass, $e$ is the electron charge, $v_{\rm F}$ is the Fermi velocity, and $n_{\rm H}$ is the Hall carrier density.

The value of this product should only be dependent on the material for good metals.
We use the value of \SI{1.4e8}{\cm\per\second} for the Fermi velocity~\cite{warburton_double_1995,bahrami_vortex_2025}, assumed to be isotropic.
Using the resistivity values reported in Fig.~\ref{fig:transport}, we report our estimates for mean free path ($\ell$) in Fig.~\ref{fig:mfp}.
We can also calculate the estimate value of the $\rho\ell$ product, and find it to be \SI{8.3e-7}{\nano\ohm\meter^{2}} across all measurements of alpha-Ta films.
We note that our value is about double the value reported once before \SI{4.2e-7}{\nano\ohm\meter^{2}}~\cite{sauerzopf_anisotropy_1987}.
The higher product from our measurements suggests that our reported values for $\ell$ could potentially be overestimated by around a factor of two, possibly caused by assumptions made for the $v_{\rm F}$.
For the purposes of this study we report the value of $\ell$ only for comparing our argon and krypton films, for which any overestimation will be consistent across all films.

\subsection{Comparison of mean free path and coherence length}
\begin{figure}
\centering
\includegraphics[scale=1]{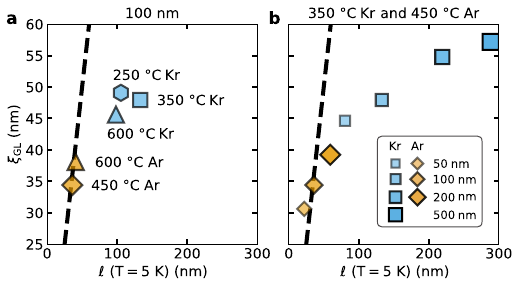}
\caption{\label{fig:clean}
Comparison of mean free path ($\ell_{\rm 5K}$) and coherence length ($\xi_{\rm GL}$). The black dashed line indicates $\ell_{\rm 5K}=\xi_{\rm GL}$; data to the right trend towards clean limit superconductivity.}
\end{figure}

In Fig.~\ref{fig:clean} we compare $\ell$ and $\xi_{\rm GL}$ for films with different deposition conditions and different thicknesses.
Both of these quantities are estimates, so they are mainly useful for relative comparison between argon and krypton films.
For every thickness, krypton-deposited films have both higher $\ell$ and $\xi_{\rm GL}$ than argon.
We can also compare the estimate for $\ell$ to thickness ($\rm d$) and to $\xi_{\rm GL}$.
Krypton-deposited films find $\ell \sim \rm d$, consistent with surface-scattering limited resistivity. 
These films also typically exhibit $\ell\gtrsim\xi_{\rm GL}$, suggesting that the superconducting state is potentially in the clean limit. 
In contrast, Ar-deposited films exhibit resistivity limited by internal scattering and are more dirty.

\begin{table}[]
    \centering
    \begin{tabular}{c||c|c|c|c|c|c|c}
       Film & \begin{tabular}{@{}c@{}}$\rho_{\rm 305K}$ \\ (\si{\micro\ohm\centi\meter})\end{tabular} & \begin{tabular}{@{}c@{}}$\rho_{\rm 5K}$ \\ (\si{\micro\ohm\centi\meter})\end{tabular} & RRR & \begin{tabular}{@{}c@{}}T\textsubscript{c} \\ (\si{\kelvin})\end{tabular} & \begin{tabular}{@{}c@{}}$\xi_{\rm GL}$ \\ (\si{\nano\meter})\end{tabular} & \begin{tabular}{@{}c@{}}$\ell_{\rm 305K}$ \\ (\si{\nano\meter})\end{tabular} & \begin{tabular}{@{}c@{}}$\ell_{\rm 5K}$ \\ (\si{\nano\meter})\end{tabular}\\
       \hline
       Bare & 16.26 & 0.53 & 30.4 & 4.32 & 47.9 & 4.80 & 133 \\
       Fabricated & 16.06 & 0.57 & 28.4 & 4.29 & 50.7 & 4.90 & 143 \\
       Resonator & 16.10 & 0.56 & 28.8 & 4.27 & 45.0 & N/A & N/A
    \end{tabular}
    \vspace*{3mm}
    \caption{Comparison of various transport measurements of \SI{100}{\nano\meter} Ta film deposited with Kr at \SI{350}{\celsius}.
    Bare are values for sample without any subsequent processing, as described in Sec.~\ref{sec:resistivity}.
    Fabricated are the same measurements done for a unpatterned portion of a film subjected to the entire resonator fab.
    Resonator are 4-point longitudinal measurements along the feed line of a patterned resonator.
    We do not have Hall measurements for the resonator sample.}
    \label{tab:res_postfab}
\end{table}

In Tab.~\ref{tab:res_postfab} we report the electronic properties of a \SI{100}{\nano\meter} Ta sample deposited at \SI{350}{\celsius} with Kr, as an unprocessed bare film, processed bare film, and an etched resonator.
We observe slight variations in most measured values, however all three samples maintain their high values for RRR, $\xi_{\rm GL}$, and $\ell_{\rm 5K}$, after etching and processing.

\subsection{X-ray diffraction}
$\theta$-2$\theta$ X-ray diffraction (XRD) scans were performed on a Rigaku Smartlab X-ray Diffractometer with a Cu K-\textalpha source to assess the phase composition.

Strong BCC \textalpha-Ta 110 peaks at about \SI{38.5}{\degree} (2$\theta$) were observed for heated films. Small deviations in the peak position likely originate from residual film strain in the samples. 
The Ta film sputtered in Kr at room temperature exhibits a weak and broadened \textalpha-Ta 110 signal along with a strong \textbeta-Ta 002 signal. This likely originate from small \textalpha-Ta crystals embedded in the predominantly \textbeta-Ta film, which were observed in TEM.

Additional peaks at approximately \SI{34.6}{\degree} and \SI{36.7}{\degree} were observed. We correlate the \SI{34.6}{\degree} to \textalpha-Ta 110 crystal diffracted with leakage Cu K-\textbeta\;rays, while the origin of the other peak is unclear. Both peaks disappear when a Ge-220 4-bounce monochromator was used, suggesting they are instrument artifacts. The monochromator, however, significantly increases data acquisition time.

Si-002 peaks have a "left shoulder" and is attributed to film-induced substrate strain. This was verified by applying a \SI{0.5}{\degree} $\omega$ offset which suppresses substrate contributions.

\begin{figure}
\centering
\includegraphics[scale=1]{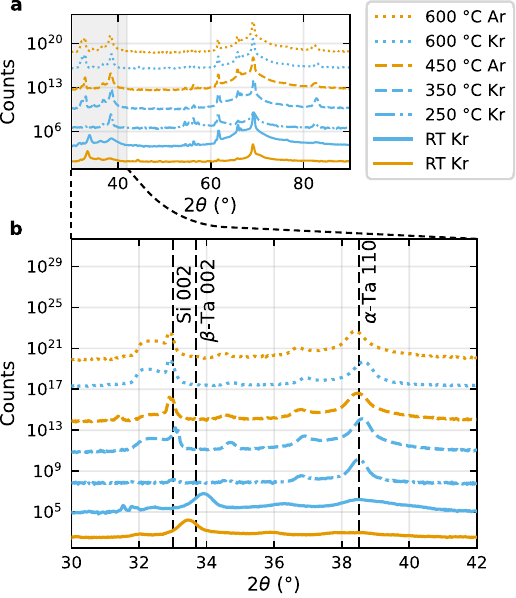} 
\caption{\label{fig:XRD} (a) Overview and (b) Zoomed-in X-ray diffraction for tantalum films sputtered at room temperature (about \SI{20}{\celsius}) to \SI{600}{\celsius}, with argon and krypton as process gases. Datasets were offset vertically for clarity. High temperatures favor formation of BCC-phase tantalum, and notably have sharper peaks indicating larger vertical grain sizes.  Deviation from expected peak angles are attributed to film stress. Overview plot (a) has been smoothened with a Savitzky-Golay filter for plotting clarity.
}
\end{figure}

\subsection{Atomic Force Microscopy}
\label{sec:afmrange}

\begin{figure}
\includegraphics[scale = 1]{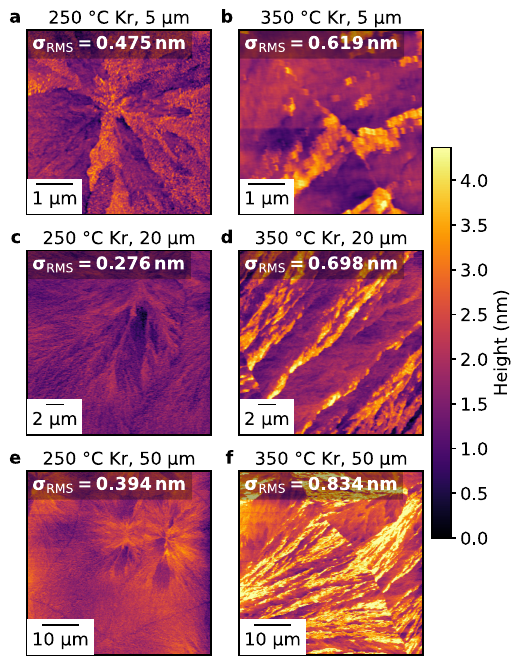}
\caption{\label{fig:afmranges} The film structure is visibly different between Ta films grown in Kr for \SI{250}{\celsius} (a, c, e), and for \SI{350}{\celsius} (b, d, f). Closer views shows bumpy areas alongside smoother areas distinctly apart, instead of being homogenously rough. Broader scans shows that the higher temperature shows streaks of features.}
\end{figure}

The growth mode switching between an intermediate temperature to a higher temperature, as observed before, can be visible when comparing the AFM images at different ranges. There is also a significant increase in roughness, as macroscopic roughness features emerge in the \SI{350}{\celsius} sample as compared to the \SI{250}{\celsius} sample.

\subsection{Electron back-scattering diffraction maps}

Electron back-scattering diffraction was performed in a Zeiss Sigma 500 SEM system using a Bruker eFlash$^\text{HR}$ detector. 
The process fits the scattered pattern from a point on the film to Kikuchi patterns. 
Films were mounted as-grown at a \SI{70}{\degree} angle with \SI{18}{\milli\meter} between the sample and the detector. 
The working distance was between \SI{16.0}{\milli\meter} to \SI{22.4}{\milli\meter}. 
Electron gun voltage was set to \SI{20}{\kilo\volt} with beam aperture set to \SI{300}{\micro\meter}. Pixel sizes of \SI{0.41}{\micro\meter} was used for \SI{200}{\celsius} and \SI{350}{\celsius} films with a range of \SI{200}{\micro\meter} , while the \SI{600}{\celsius} film used a \SI{1.02}{\micro\meter} pixel size with \SI{500}{\micro\meter} range. 
The wider range captures the larger features observed for the \SI{600}{\celsius} film. 

\begin{figure}
\centering
\includegraphics[scale=1]{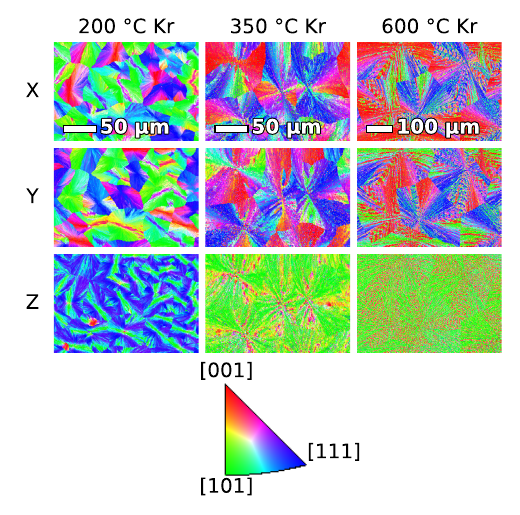}
\caption{\label{fig:EBSD} Electron back-scattering diffraction analysis for film sputtered in Kr plasma at different temperatures. For a clearer visualization, these EBSD images have been post processed. Pixels with no fits were recursively filled with the average color of the nearest neighboring colored pixels. Kikuchi patterns were well for \SI{95.5}{\percent}, \SI{51.9}{\percent}, and \SI{71.4}{\percent} of the pixels for the \SI{200}{\celsius}, \SI{350}{\celsius}, and \SI{600}{\celsius} respectively. While change is not apparently from the X and Y orientations, the Z orientation shows a transition between lower temperature to higher temperature, indicative of some kind of phase transition in sputtered growth as temperature increased.}
\end{figure}

The EBSD maps reflects boundaries similar to those observed on the AFM. \SI{350}{\celsius} and \SI{600}{\celsius} films shows a phase shift in the crystal orientations and grain structure as compared to the \SI{200}{\celsius} films. 
In these higher temperature films, a shift in crystal orientation can be observed in the Z axis EBSD maps, and streaks appear in the X and Y axis EBSD maps. 
The difference in Z axis maps suggests a transition in the growth process has occurred. 
This supports the AFM-based observation in a change in growth modes between \SI{200}{\celsius} and \SI{350}{\celsius}. 
The streaks in the X and Y axis suggest the formation of micro-grains of differently oriented crystals. 
This possibility is also supported by the fact that Kikuchi pattern fitting on the \SI{200}{\celsius} film produced noticeably better fits than the two higher temperature films.

The \SI{200}{\celsius} film sampled in the EBSD looked similar to the \SI{250}{\celsius} film under atomic force microscopy (AFM), with transport measurement and XRD data characteristic of \textalpha-Ta. Thus we believe the EBSD map is also representative of \SI{250}{\celsius} films mentioned in other parts of the paper.

\subsection{Scanning Transmission Electron Microscopy}
\begin{figure*}[t]
\includegraphics[width=\textwidth]{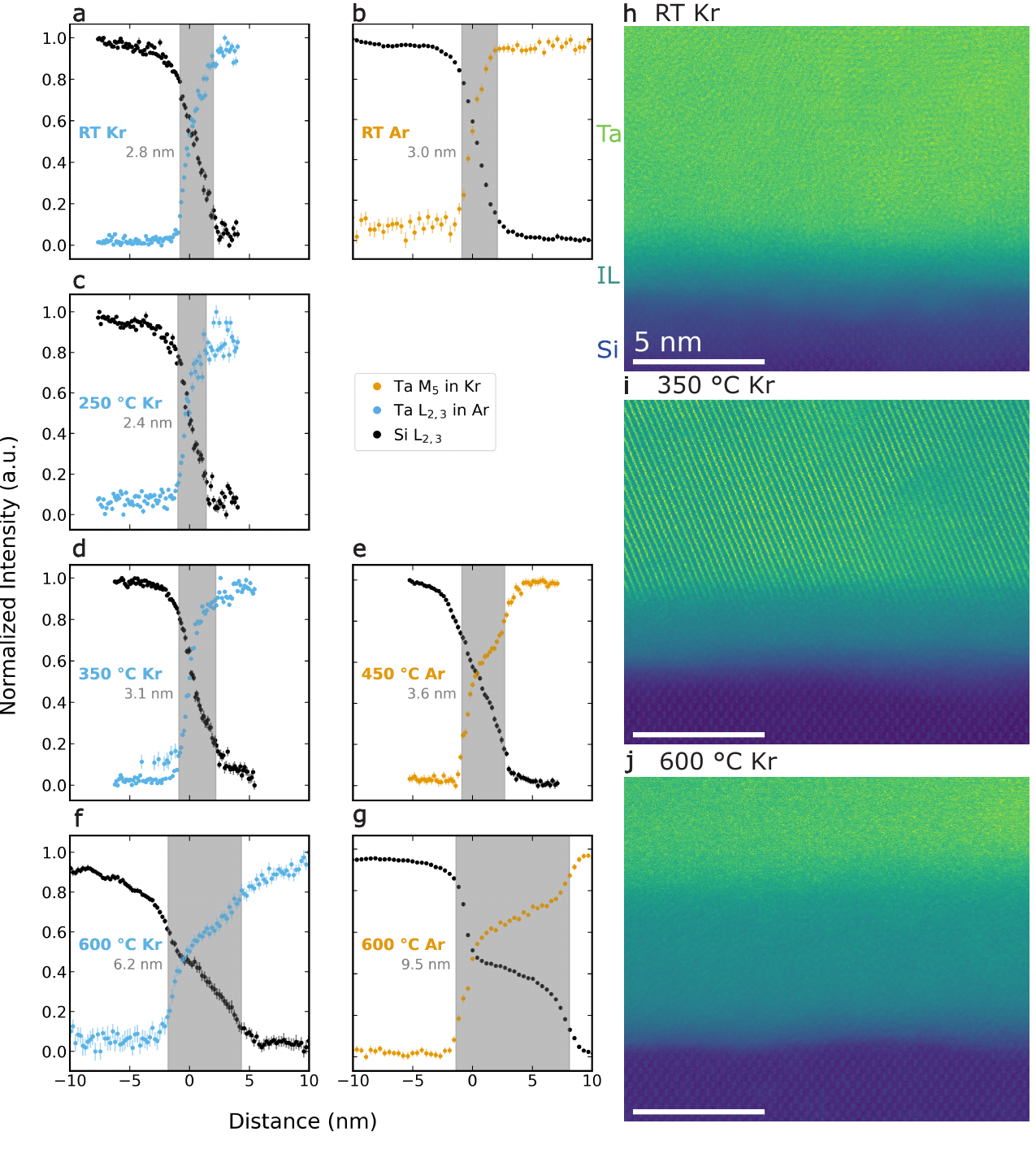}%
\caption{\label{fig:eels} (a-g) EELS line profiles across the metal-substrate interface for 100 nm Ta films under deposition conditions indicated in the figure panel.
(h,i,j) False-colored ADF-STEM images of the interface for Kr-deposited samples at three temperatures.
}
\end{figure*}

For specimen preparation we used a chemically-inert Xe plasma in the focused ion beam (FIB) rather than Ga, given the tendency of Ga to react with Ta and introduce artifacts, especially at surfaces and interfaces. Inverted cross-sectional lamellas of the deposited films, before device fabrication, were prepared on a TESCAN Amber X2 plasma FIB-SEM with low-voltage polishing final steps following typical focused-ion-beam lift-out procedures.  ADF-STEM images were acquired at 300 kV and a probe convergence angle of 30 mrad on aberration-corrected Thermo Fisher Scientific (TFS) Spectra 300 X-CFEG, while EELS data was collected at 120 kV, 21.4 mrad convergence angle, and a camera length of 33 mm. The normalized counts of the $\mathrm{Si}\ \mathrm{L}_{2,3}$ and $\mathrm{Ta}\ \mathrm{M}_{5}$ edges were used to generate a composition profile of the metal-substrate interface along the growth direction of the films (Fig.~\ref{fig:eels}).

Consistent with the ADF-STEM images, Fig.~\ref{fig:eels}(a) shows that elevated substrate temperatures during deposition significantly increase the thickness of the Ta-Si interlayer. In samples grown at \SI{450}{\celsius} and above, the distinctive shoulder in the Ta profile indicates the presence of a silicide, further confirmed by a transformation in the shape of the $\mathrm{Si}\ \mathrm{L}_{2,3}$ edge within the interlayer.  Probing the interface for the oxygen K edge additionally confirms the lack of oxygen at the metal-substrate interface.

\subsection{Secondary Ion Mass Spectrometry} \label{app:SIMS}
\begin{figure}
\centering
\includegraphics[width=\linewidth]{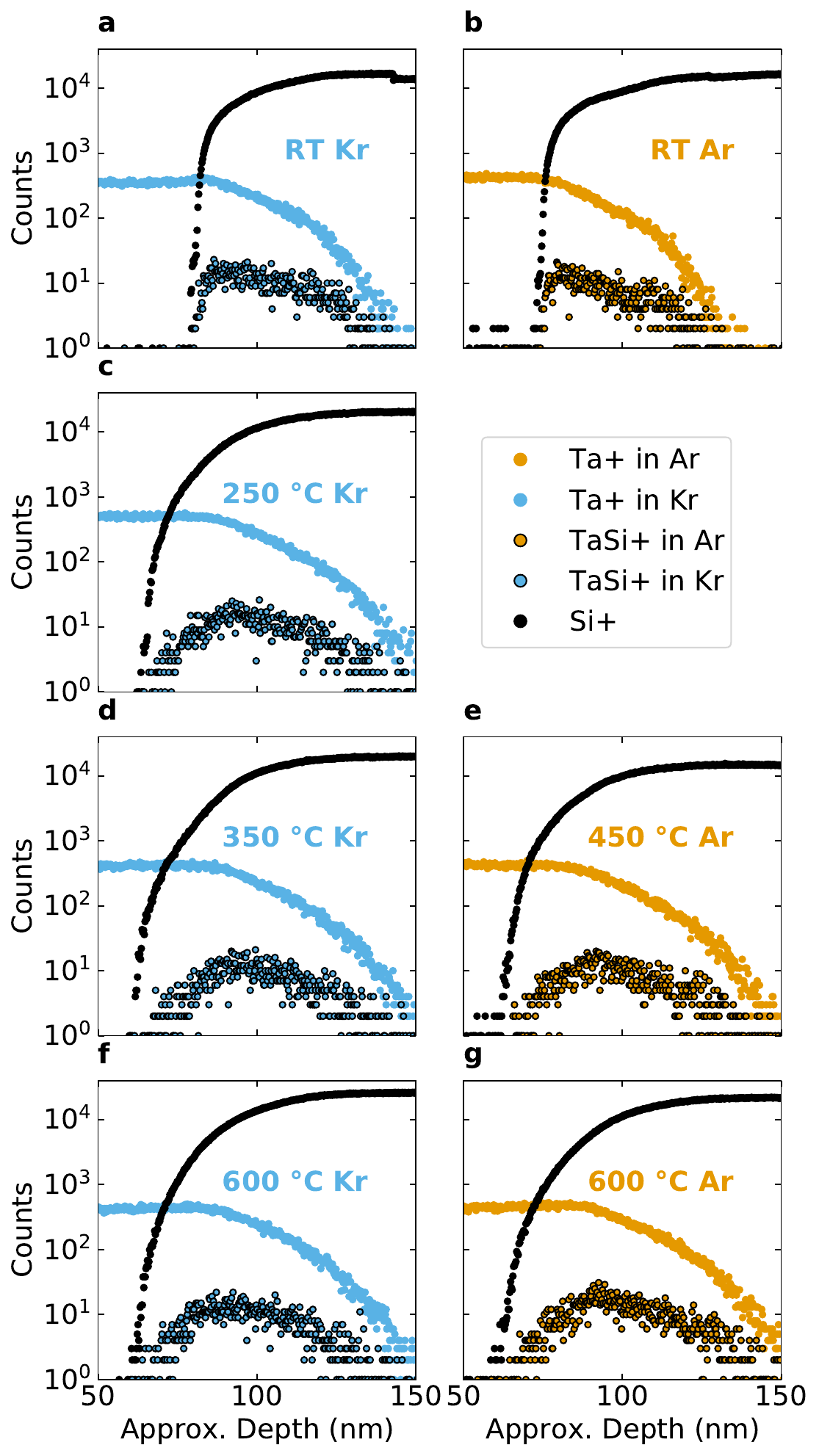}
\caption{\label{fig:sims}
SIMS data for \SI{100}{\nano\meter} Ta films deposited with both Kr and Ar at various temperatures, profiled using oxygen sputter ions.
Selected raw counts of Ta+, Si+, and TaSi+ are plotted.
}
\end{figure}

TOF-SIMS data was collected using a TOF-SIMS V (ION-TOF USA Inc, Chestnut Ridge, NY), equipped with a \SI{25}{\kilo\electronvolt} Bismuth liquid metal ion gun (LMIG), and low energy Oxygen and Cesium ion sources.
Spectra were acquired with high mass resolution using LMIG in  ‘Spectrometry’ setting, previously called high current bunched (HCBU). Dynamic SIMS consisted of dual-beam operation, by alternating stages of analysis (Bi primary ions) and material removal (O\textsubscript{2} or Cs ions), until reaching substrate.

For best sensitivity to metal impurities (metal interfacial species), the O\textsubscript{2} sputter source was used, and for improved yield of noble gases, additional profiles were collected with Cs source (Ar, Kr, as clusters with Cs, M-Cs positive mode).

Oxygen depth profiles were conducted with \SI{2}{\kilo\electronvolt} sputter beam (\SI{780}{\nano\ampere} current); analysis area of \SI{500}{\micro\meter\squared}, was centered inside \SI{800}{\micro\meter\squared} erosion crater.
Depth of \SI{100}{\nano\meter} was approximated at the half maximum of the Si counts (in Si substrate).

Cesium profiles utilized \SI{1}{\kilo\electronvolt} energy (\SI{105}{\nano\ampere} current), and depth profiles consisted of \SI{250}{\micro\meter\squared} analysis area, centered inside \SI{800}{\micro\meter\squared} sputter crater.
Depth of \SI{100}{\nano\meter} was estimated at the point at the half maximum of the CsTa counts (in Ta film).

The SIMS plots in Fig.~\ref{fig:sims} obtained with an O\textsubscript{2} source include only select ions (Ta+, Si+, and TaSi+) observed to highlight the main differences found in the films.
We focus on the substrate-metal interface to highlight qualitative differences in Ta-Si intermixing.
We see that the overlap for Ta+ and Si+ is much smaller for RT samples, which could indicate reduced intermixing, but it is not the only possible explanation.
Qualitatively, the TaSi+ signal in the RT and the heated samples is different, with counts reaching a higher maximum value for the RT samples, while being spread out over a narrower range.
We also remark on the broader transition from majority-Ta to majority-Si in the higher-temperature depositions. 
This is likely due to inhomogeneous etching resulting from the large-area grains, which was observed during the SIMS analysis. 
Such inhomogeneous etching was also observed in the etching used in the resonator and transmon fabrication as discussed in Sec.~\ref{sec:roughness}.

\begin{figure}
\centering
\includegraphics[width=\linewidth]{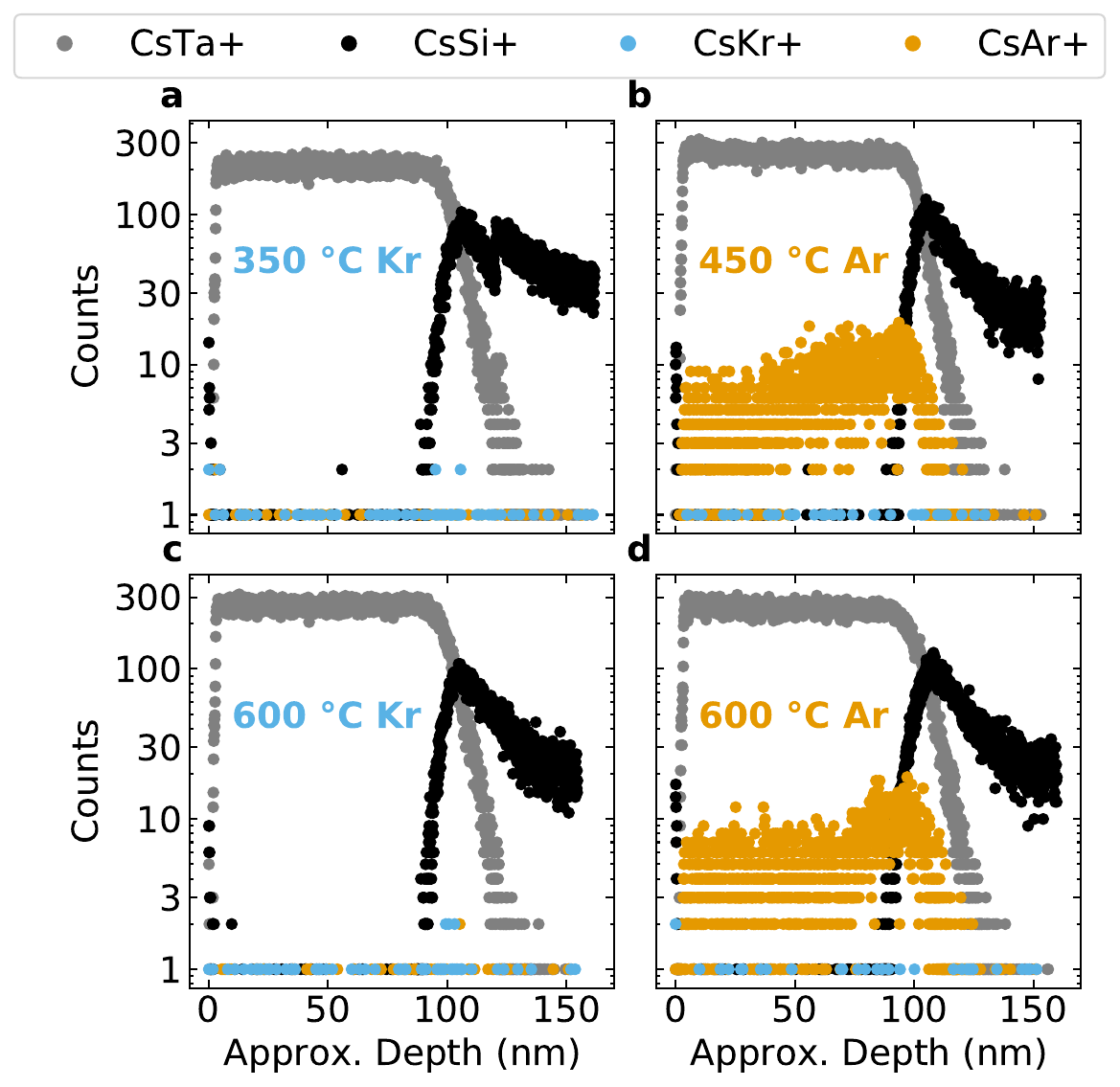}
\caption{\label{fig:sims_gas}
SIMS data for \SI{100}{\nano\meter} Ta films deposited with both Kr and Ar at various temperatures, profiled using cesium sputter ions.
Selected raw counts of CsTa+, CsSi+, CsAr+ and CsKr+ are plotted.
}
\end{figure}

SIMS spectra using the Cs source are shown in Fig.~\ref{fig:sims_gas}.
The films sputtered with Kr do not show any detectable signal for both Kr and Ar, while the samples sputtered with Ar have a clear Ar signal.
Different detection rates and yields can account for differences in Ar and Kr signal, which prevents us from doing a quantitative analysis.
It has been previously reported than when different noble gases are used as sputter gas, the gas concentration implanted (as impurity in metal film) will vary~\cite{benvenuti_study_1999}.
Different impurity levels of Ar and Kr could contribute to different differences in nucleation sites, resistivity, and crystallinity, as reported.

\subsection{Substrate roughness post etching}
\label{sec:roughness}

\begin{figure}[t]
\includegraphics[scale=1]{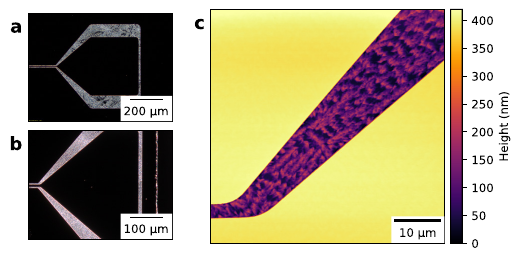}
\caption{\label{fig:postetch} Roughness features on the silicon surface of a) \SI{350}{\celsius} and b) \SI{600}{\celsius} tantalum films are clearly visible in dark field optical microscopy. c) AFM of etched \SI{600}{\celsius} tantalum film, showing clearer the roughness. In this image, RMS roughness is \SI{1.7}{\nano\meter} and \SI{48}{\nano\meter} for Ta and Si surfaces respectively.}
\end{figure}

The etched Si surface of the high-temperature deposited Ta films was found to be significantly rougher than that of the room-temperature deposited films. 
Our Ta etch is a low-power, highly-chemical, ICP-RIE etch with a \SI{12}{\watt} (\SI{60}{\volt}) RIE bias, and \SI{800}{\watt} ICP power.
As shown in Fig.~\ref{fig:postetch}, the large grain structure of the heated Ta films is imprinted on the surface of the Si after the etch.
This effect is also visible in the images in Fig.~\ref{fig:resonator}(a) and Fig.~\ref{fig:transmon}(a).
This suggests crystal-orientation-dependent etch rates of the tantalum films.

\section{Additional Resonator Measurements}

\begin{figure}[t]
\includegraphics[scale=1]{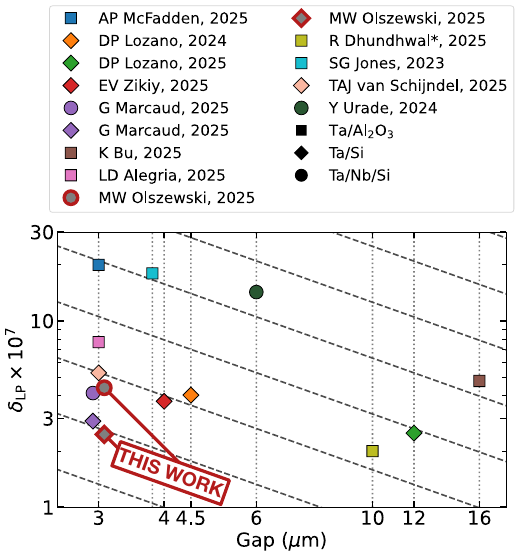}
\caption{\label{fig:literature} 
Comparison plot for losses with literature~\cite{mcfadden_fabrication_2024,lozano_low-loss_2024,lozano_reversing_2025,zikiy_investigation_2025,marcaud_low-loss_2025,dhundhwal_high_2025,jones_grain_2023,van_schijndel_cryogenic_2025,urade_microwave_2024,bu_tantalum_2025,alegria_growth_2025}.
Data points with \SI{3}{\micro\meter} gap are offset for clarity.
}
\end{figure}

The summary of all of the resonators measured is presented in Tab.~\ref{tab:resonators}.
Resonator comparison with literature is presented in Fig.~\ref{fig:literature}.

\begin{table}
    \centering
    \begin{tabular}{c|c|c|c|c|c|c||c|c|c}
        \multicolumn{1}{c}{}&\multicolumn{1}{c}{}&\multicolumn{1}{c}{} &\multicolumn{1}{c}{} &\multicolumn{1}{c}{} &\multicolumn{1}{c}{} & & \multicolumn{3}{c}{$\delta_{\rm LP}\times10^{7}$} \\
        Gas & Temp & Seed & BOE & Etch & N\textsubscript{c} & N\textsubscript{p} & med & iqr & std \\
        \hline
        \hline
        Ar & RT & Nb & N & S & 1 & 4 & 9.70 & 2.85 & 1.77 \\
        Ar & RT & Nb & Y & S & 3 & 17 & 4.78 & 1.99 & 1.34 \\
        Ar & RT & Nb & A & S & 1 & 5 & 9.80 & 0.22 & 1.62 \\
        \hline
        Kr & 250 & N & Y & S & 2 & 16 & 3.10 & 1.24 & 1.30 \\
        Kr & 250 & N & A & S & 1 & 8 & 5.63 & 1.27 & 0.93 \\
        Kr & 350 & N & N & S & 2 & 15 & 7.36 & 2.51 & 3.20 \\
        Kr & 350 & N & Y & S & 2 & 15 & 2.45 & 0.80 & 0.97 \\
        Kr & 350 & N & A & S & 1 & 8 & 6.62 & 1.06 & 1.40 \\
        Kr & 350 & N & Y & D & 2 & 15 & 2.79 & 0.82 & 0.90 \\
        Ar & 450 & N & Y & S & 2 & 16 & 2.84 & 1.36 & 0.97 \\
        Ar & 600 & N & N & S & 1 & 5 & 11.36 & 3.10 & 3.00 \\
        Ar & 600 & N & Y & S & 1 & 1 & 12.5 & N/A & N/A \\
        Kr & 600 & N & N & S & 1 & 8 & 8.17 & 1.87 & 3.29 \\
        Kr & 600 & N & Y & S & 2 & 15 & 8.19 & 3.68 & 3.57 \\
    \end{tabular}
    \vspace*{3mm}
    \caption{\label{tab:resonators}
    Summary of all of the resonators measured for the this study.
    Gas represents the process gas used during the deposition.
    Temp denotes the substrate temperature during the deposition.
    Seed distinguishes samples with and without a \SI{5}{\nano\meter} seed layer.
    BOE marks samples with a post-BOE treatment prior to loading for measurements with no (N), yes (Y), and samples with post-BOE and at least 5 weeks of aging (A).
    Etch signifies if the over etch into the silicon substrate is shallow (S) or deep (D).
    N\textsubscript{c} is the number of chips measured.
    N\textsubscript{p} is the number of resonators measured at single photon powers that properly fitted to obtain $\delta_{\rm LP}$ value.
    The right side of the table shows the statistics for $\delta_{\rm LP}$ including median (med), interquartile ranges (iqr), and standard deviations (std).
    }
\end{table}

\subsection{Acid-Free \textalpha-Ta resonators}\label{app:acid_free}

\begin{figure*}
\centering
\includegraphics[width=0.9\textwidth]{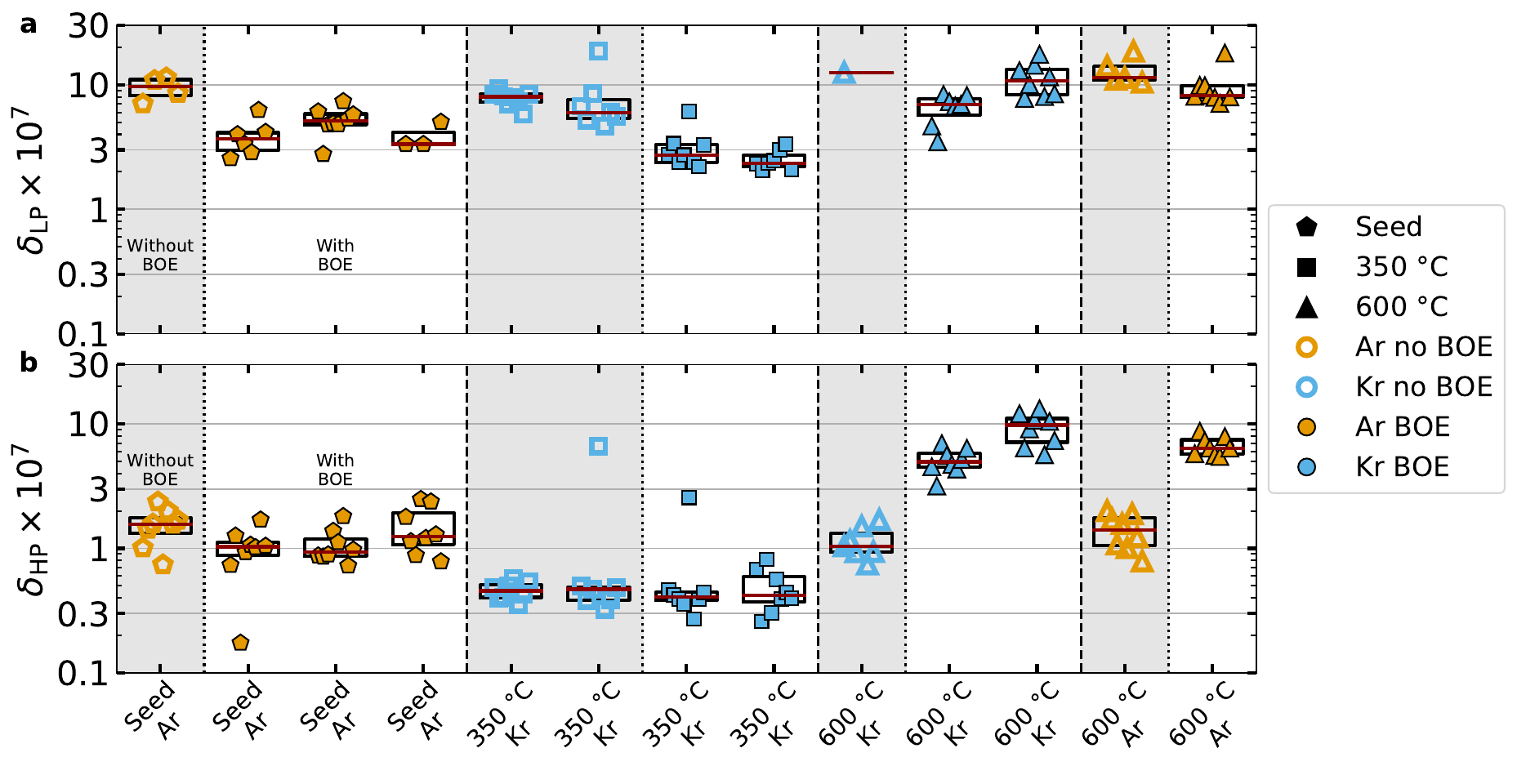}
\caption{\label{fig:noBOE}
(a) Low power performance of resonators deposited with the same conditions without and with a post-BOE treatment.
Grey areas highlight samples without post-BOE treatments.
(b) High power performance of the same films.
}
\end{figure*}

Resonator performance can be significantly improved with various post-fabrication treatments which remove native tantalum and silicon oxides~\cite{urade_microwave_2024,crowley_disentangling_2023}.
Buffered oxide etchant (BOE) is commonly used for this purpose.
As previously reported~\cite{olszewski_low-loss_2025}, our resist cleaning procedure using AZ300T results in clean surfaces based on X-ray photoelectron spectroscopy analysis. 
Nb resonators made this way exhibited high quality factors without BOE treatment.
In Fig.~\ref{fig:noBOE}, we report resonator performance without BOE treatment.
The losses in these devices were higher than the losses in the BOE-treated devices.

\subsection{Aged \textalpha-Ta resonators}\label{app:aging}
\begin{figure}
\centering
\includegraphics[width=\linewidth]{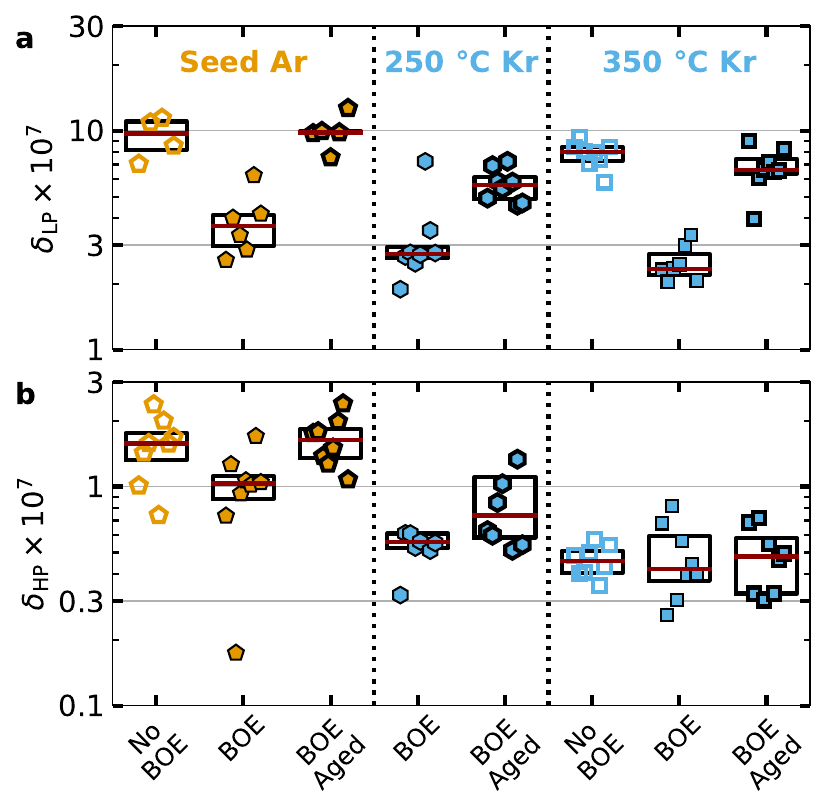}
\caption{\label{fig:aging}
(a) Low power performance of Nb-seeded Ar RT, \SI{250}{\celsius} Kr, and \SI{350}{\celsius} \SI{100}{\nano\meter} Ta films.
No BOE samples did not have a post-BOE treatment prior to loading into the cryostat.
BOE samples had a \SI{1}{\minute} treatment in 10:1 BOE.
BOE aged samples are the same devices as the BOE samples, but unloaded and aged at atmospheric conditions (without removing from the packages) for at least 5 weeks.
(b) High power performance of the same films.
}
\end{figure}

Post-processing that reduces native surface oxides is frequently observed to significantly improve resonator performance~\cite{verjauw_investigation_2021,altoe_localization_2022,zheng_nitrogen_2022,crowley_disentangling_2023,lozano_low-loss_2024,lozano_reversing_2025}.
Furthermore, studies have shown that the aging of Nb resonators in ambient conditions results in increased losses, correlated with regrowth of NbO\textsubscript{x}~\cite{altoe_localization_2022,verjauw_investigation_2021}.

Here, we show that Nb-seeded Ta resonators age differently Ta resonators deposited directly on Si by comparing with Kr-deposited films at \SI{250}{\celsius} and \SI{350}{\celsius}. 
In Fig.~\ref{fig:aging}, we present aging data of resonators treated with BOE prior to measurements, which were then left in atmospheric conditions for at least 5 weeks, and measured again to assess the aging of our devices.
The BOE treatment is \SI{1}{\minute} in 10:1 BOE, which removes the native silicon oxide and partially removes the niobium oxide~\cite{verjauw_investigation_2021,altoe_localization_2022,olszewski_low-loss_2025}.
The chips were never removed from their measurement package, so that the wire bonding configuration is the same.
We observe that aging of the Nb-seeded resonators results in increases of both the LP and HP losses, whereas aging the Ta-on-Si resonators only results in increased LP losses. 
In absolute terms, the Ta-on-Si resonators remain lower loss after aging.

\subsection{Deep Etching of Resonators}\label{app:deep_etch}
\begin{figure}
\centering
\includegraphics[width=\linewidth]{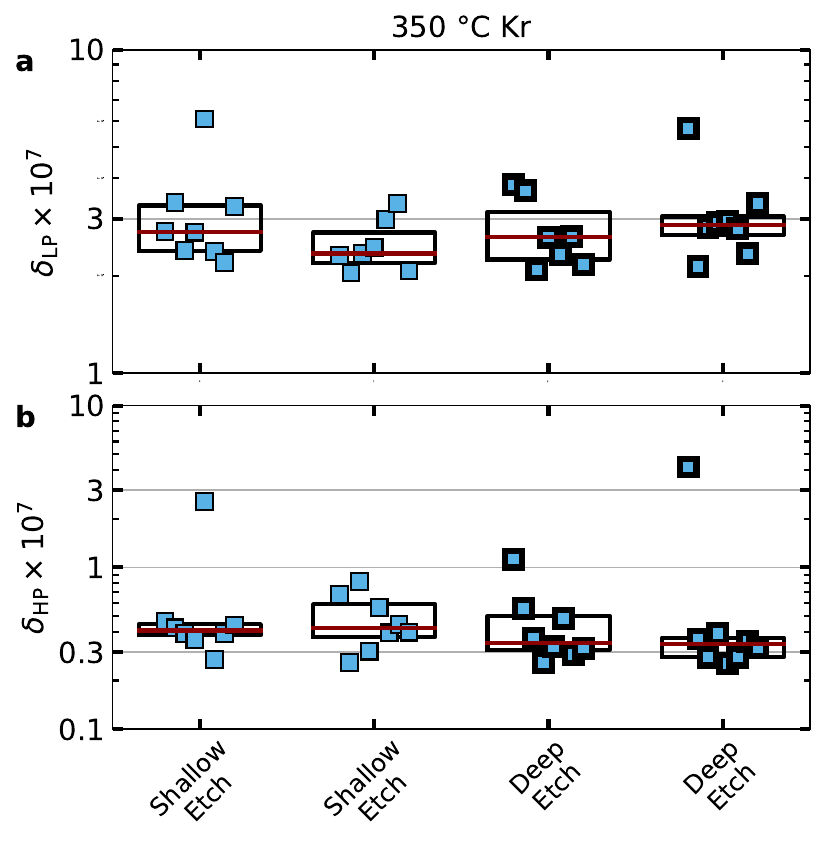}
\caption{\label{fig:deep}
(a) Low power performance of \SI{350}{\celsius} Kr resonators with a shallow and a deep over etch into the silicon substrate.
Stars indicate median HP loss.
(b) High power performance of the same films.
The data for the shallow etch samples is the same as in the main text.
}
\end{figure}

A recent report found that over-etching into the substrate can improve resonator quality factor~\cite{bruno_reducing_2015,zikiy_investigation_2025}.
Our attempt at this, over-etching by around 500-700\si{\nano\meter}, did not meaningfully lower resonator losses.
The comparison with films deposited at \SI{350}{\celsius} with Kr is shown in Fig.~\ref{fig:deep}.
Our etch chemistry is different than that reported by Zikiy, \textit{et al.}~\cite{zikiy_investigation_2025}, which may explain the discrepancy.

\section{Additional Transmon Measurements}\label{app:transmons}
\begin{table*}
    \centering
    \begin{tabular}{c|c|c|c||c|c|c||c|c|c||c|c|c}
        \multicolumn{1}{c}{} &\multicolumn{1}{c}{} &\multicolumn{1}{c}{} & &\multicolumn{3}{c}{$\rm Q_{\rm 1}~(10^{7})$} & \multicolumn{3}{c}{T\textsubscript{2R} (\si{\micro\second})} & \multicolumn{3}{c}{T\textsubscript{2H} (\si{\micro\second})}\\
        Number & Descum & E\textsubscript{J}/E\textsubscript{C} & $f_{q}$ (\si{\mega\hertz}) & med & iqr & std & med & iqr & std & med & iqr & std \\
        \hline
        \hline
        1 & N & N/A & 2461 & 9.05 & 3.1 & 2.12 & 19 & 27 & 19 & 81 & 11 & 9 \\
        2 & N & 40.8 & 3169 & 14.0 & 3.3 & 2.4 & 10 & 16 & 11 & 138 & 24 & 17 \\
        3 & N & 35.4 & 3134 & 14.4 & 3.4 & 2.5 & 17 & 4 & 4 & 80 & 23 & 15 \\
        \hline
        4 & Y & 55.5 & 3812 & 6.1 & 1.6 & 1.3 & 85 & 67 & 40 & 273 & 34 & 27 \\
        5 & Y & 66.9 & 3834 & 2.6 & 0.6 & 0.4 & 32 & 14 & 11 & 92 & 45 & 28 \\
        6 & Y & 55.3 & 3801 & 7.8 & 1.4 & 1.2 & 149 & 60 & 36 & 307 & 57 & 41 \\
        7 & Y & 67.9 & 3880 & 7.3 & 2.7 & 1.9 & 109 & 46 & 35 & 215 & 63 & 40 \\
    \end{tabular}
    \vspace*{3mm}
    \caption{\label{tab:transmons}
    Summary of transmons measured for this study.
    Descum indicates if an oxygen descum was done before JJ deposition no (N) and yes (Y).
    E\textsubscript{J}/E\textsubscript{C} is the ratio of the Josephson and charging energies.
    $f_{q}$ is the qubit frequency.
    Q\textsubscript{1}, T\textsubscript{2R}, and T\textsubscript{2H} are the quality factor, Ramsey coherence time, and Hahn echo coherence times, respectively.
    Median (med), interquartile range (iqr), and standard deviation (std) are given. Device 1 has much lower anharmonicity and dispersive shift than expected from design, suggestive of two junctions in series. For this device we therefore cannot extract $E\textsubscript{J}/E\textsubscript{C}$ from the measured anharmonicity and qubit frequency.
    }
\end{table*}

\begin{figure*}
    \centering
    \includegraphics[width=0.9\textwidth]{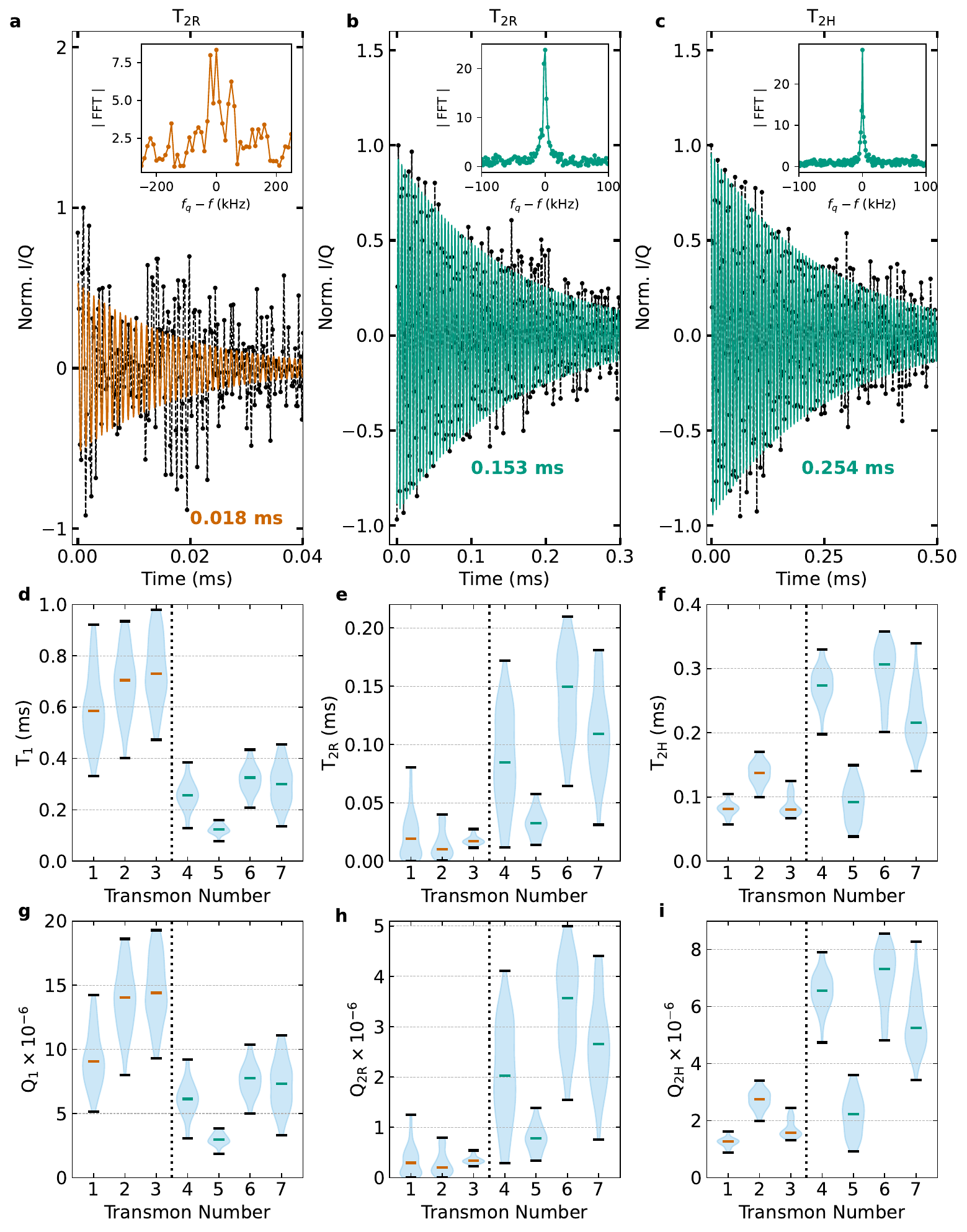}
    \caption{
    Details of transmon measurements.
    (a) Sample Ramsey coherence time (T\textsubscript{2R}) near the median value for the no-descum samples.
    Y-axis plots normalized I/Q signal.
    Plot shows significant beating which results in unreliable fitting for T\textsubscript{2R} times.
    Insert shows Fourier spectrum of the plot, with multiple noticible peaks.
    (b) T\textsubscript{2R} near the median value for the descum samples.
    Y-axis plots normalized I/Q signal.
    Insert shows fast Fourier transform (FFT), with a single peak at the qubit frequency ($f_{q}$). 
    Beating effects are not observed, unlike in subplot (a).
    (c) T\textsubscript{2H} near median value for the descum samples.
    Y-axis plots normalized I/Q signal.
    T\textsubscript{2H} for the no-descum samples did not observe the beating issue, as shown in subplot (a).
    Insert shows the Fourier spectrum.
    (d-f) Violin plots for T\textsubscript{1}, T\textsubscript{2R}, and T\textsubscript{2H}, respectively.
    All seven qubits are shown.
    (g-i) Violin plots for $\rm Q_{1}=2\pi \rm T_{1}f_{q}$, $\rm Q_{\rm 2R}=2\pi \rm T_{2R}f_{q}$, and $\rm Q_{\rm 2H}=2\pi \rm T_{2H}f_{q}$, respectively.
    All seven qubits are shown.
    }
    \label{fig:transmon_t2}
\end{figure*}

Complete results of transmon measurements are reported in Tab.~\ref{tab:transmons} and Fig.~\ref{fig:transmon_t2}.

First, we highlight the significant beating observed in Ramsey coherence time (T\textsubscript{2R}) measurements, as shown in  Fig.~\ref{fig:transmon_t2}a.
The example fit is representative median device performance for samples without an oxygen descum.
This effect is most prevalent, but not limited to, transmons prepared without a descum treatment, resulting in low confidence of the fits for T\textsubscript{2R} for these samples.
In contrast, transmons prepared with an oxygen descum typically do not display this issue, as shown in Fig.~\ref{fig:transmon_t2}b.
As highlighted by the Fourier transmon of both plots, the beating is a result of low-frequency noise, indicated by multiple peaks, as compared to one peak.
Low frequency noise can be partially eliminated with Hahn-echo coherence time  (T\textsubscript{2H}) measurement, 
Our T\textsubscript{2H} data do not show significant beating for both sets of transmons, as illustrated in Fig.~\ref{fig:transmon_t2}c.

Fig.~\ref{fig:transmon_t2}d and g includes the same data as shown in the main text Fig.~\ref{fig:transmon}, highlighting the differences in T\textsubscript{1} and Q\textsubscript{1}, respectively, between the two sets of transmons.
Similarly, in Fig.~\ref{fig:transmon_t2}e and h, we present data for T\textsubscript{2R} and Q\textsubscript{2R} ($2\pi \rm T_{\rm 2R}f_{q}$), respectively.
Fig.~\ref{fig:transmon_t2}f and i, contains data for T\textsubscript{2H} and Q\textsubscript{2H} ($2\pi \rm T_{\rm 2H}f_{q}$), respectively.
We highlight that although the samples without the descum have much higher Q\textsubscript{1} than transmons with descum, the resulting T\textsubscript{2R} and T\textsubscript{2H} are much lower.

The explanation for this phenomenon is not fully known as this time.
Tab.~\ref{tab:transmons} highlights the observed differences in samples without and with the oxygen descum.
The ratio of Josephson energy to the charging energy E\textsubscript{J}/E\textsubscript{C} is higher for samples treated with the oxygen descum.
Additionally, the measured qubit frequency $f_{q}$ is lower for the samples without the oxygen descum.
These changes are consistent with a decrease in JJ resistance after the introduction of the oxygen descum, which itself could impact the observed performance.
We leave the investigation of the differences in transmon performance both in terms of T\textsubscript{1}, T\textsubscript{2R}, and T\textsubscript{2H}, to future work.

\section{Microwave Measurement Details}\label{app:measurement}

\subsection{Resonator Measurements}
A Copper Mountain M5180 VNA with 0 to \SI{-50}{\deci\bel m} dynamic range was used along with one programmable attenuator with tunability between 0 to \SI{60}{\deci\bel} attenuation at room temperature.
The output lines host one high electron mobility transitor (HEMT) amplifier of \SI{37}{\deci\bel} gain at \SI{4}{\kelvin}, and one amplifier at room temperature of \SI{38}{\deci\bel} gain.
The even distribution of measured points around the circle was achieved by using the homophasal point distribution method as described in~\cite{baity_circle_2024}.

\subsection{Transmon Measurements}
All measurements of transmon qubits were performed with AMD/Xilinx ZCU216 and 4x2 RFSoC boards using the open-source QICK software/firmware stack \cite{stefanazzi_qick_2021, ding_experimental_2024}. 
All signals are generated with direct RF sampling, without using analog mixers.
In this work all measurements are performed without parametric amplifiers. The wiring setup for qubit measurements is the same as shown in Fig.~\ref{fig:wiring-diagram}, using a single line for qubit drive and readout pulses. All qubit drive pulses are rectangular with a duration of $\SI{4}{\mu s}$. Readout is performed using a rectangular pulse with duration of $\SI{10}{\mu s}$. Before each shot we ensure that the qubit is in its ground state using a waiting time of at least $5\,T_1$. For each measurement of $T_1$, $T_{2R}$, and $T_{2H}$ we average over at least 300 rounds of single-shot measurements in order to minimize the fitting error for each individual data point. For each qubit we perform measurements of $T_1$, $T_{2R}$, and $T_{2H}$ over a duration of at least 12 hours in order to average over temporal fluctuations of qubit performance.

\subsection{Cryogenic Microwave Wiring Diagram}
\label{app:wiring}
\begin{figure}
\includegraphics[width=\columnwidth]{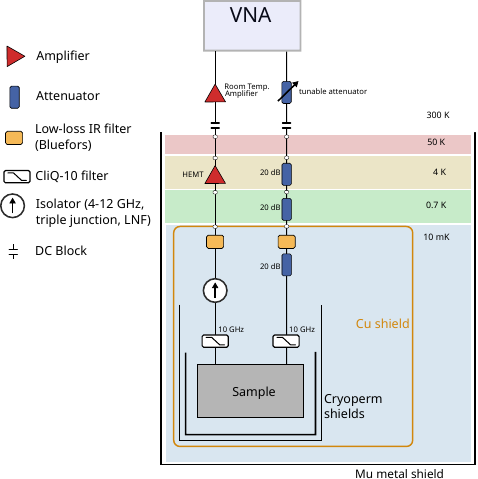}
\vspace*{3mm}
\caption{\label{fig:wiring-diagram} Schematic of cryogenic measurement setup used for resonators. For qubit measurements we replace the VNA with RFSoC boards as explained in the text.}
\end{figure}

All measurements are performed in a dilution refrigerator with base temperature below $\SI{10}{mK}$.
Samples are mounted in a coldfinger (Linqer16, Scalinq) with two layers of cryogenic magnetic shielding and a room-temperature mu-metal shield.
All microwave lines are filtered using low-loss IR filters and hybrid low-pass/IR filters with a cut-off frequency of \SI{10}{GHz}.
The details of the setup are shown in Fig.~\ref{fig:wiring-diagram}.

\clearpage

\bibliography{vf-references.bib}

\end{document}